\documentclass{aa}  

\def \eg {e.g.}

\def \ie {i.e.}

\def \omegam {{\hbox{$\Omega_{\rm m}$}}}
\def \omegal {{\hbox{$\Omega_\Lambda$}}}
\def \hzero {{\hbox{$H_0$}}}

\def \arcsec {\hbox{$^{\prime\prime}$}}
\def \deg {\hbox{$^\circ$}}
\def \nh {\hbox{$N_{\rm H}$}}

\def \msun {\hbox{${\rm M_\odot}$}}

\def \mfive {\hbox{$M_{500}$}}

\def \rfive {\hbox{$r_{500}$}}

\newcommand{\alfvenic }{Alfv\'{e}nic}

\newcommand{\ergs }{\mbox{erg s$^{-1}$}}

\newcommand{\kmsmpc }{\mbox{km s$^{-1}$ Mpc$^{-1}$}}

\newcommand{\kev }{\mbox{keV}}

\newcommand{\mujyb }{\mbox{$\mu$Jy beam$^{-1}$}}

\newcommand{\muG }{\mbox{$\mu$G}}

\newcommand{\acis }{ACIS}
\newcommand{\acisE }{Advanced CCD Imaging Spectrometer}

\newcommand{\uv }{\textit{uv}}

\newcommand{\wsclean }{\textsc{WSClean}}

\newcommand{\spam }{\textsc{spam}}
\newcommand{\spamE }{Source Peeling and Atmospheric Modeling}

\newcommand{\xspec }{\textsc{xspec}}

\newcommand{\ciao }{\textsc{ciao}}

\newcommand{\brats }{BRATS}
\newcommand{\bratsE }{Broadband Radio Astronomy Tools}

\newcommand{\xmm }{{\em XMM-Newton}}
\newcommand{\chandra }{{\em Chandra}}

\newcommand{\srgE }{Spectrum R\"{o}ntgen Gamma}
\newcommand{\erosita }{{eROSITA}}
\newcommand{\erositaE }{Extended R\"{O}ntgen Survey with an Imaging Telescope Array}

\newcommand{\ugmrt }{uGMRT}
\newcommand{\ugmrtE }{upgraded Giant Metrewave Radio Telescope}

\newcommand{\lofar }{LOFAR}
\newcommand{\lofarE }{LOw Frequency ARray}

\newcommand{\lotss }{LoTSS}
\newcommand{\lotssE }{LOFAR Two-meter Sky Survey}
\newcommand{\lolss }{LoLSS}
\newcommand{\lolssE }{LOFAR LBA Sky Survey}

\newcommand{\tgss }{TGSS}
\newcommand{\tgssE }{TIFR GMRT Sky Survey}

\newcommand{\vlss }{VLSS}
\newcommand{\vlssE }{VLA Low-frequency Sky Survey}

\newcommand{\ukidss }{UKIDSS}
\newcommand{\ukidssE }{United Kingdom Infrared Deep Sky Survey}

\usepackage{graphicx}
\usepackage{amssymb}
\usepackage{multirow,bigdelim}
\usepackage{txfonts}
\usepackage[export]{adjustbox}
\usepackage{hyperref}
\hypersetup{colorlinks,citecolor=blue,filecolor=black,linkcolor=red,urlcolor=black}
\newcommand{\sou}{VLSS J0318.9+5755}
\newcommand{\treble}{Treble Clef}

\begin{document} 

\title{The Treble Clef radio phoenix and its old nonthermal filaments} 

\authorrunning{A. Botteon et al.} 
\titlerunning{The Treble Clef radio phoenix}

\author{A. Botteon\inst{\ref{ira}},
M. Brienza\inst{\ref{ira}},
K. Rajpurohit\inst{\ref{cfa}},
N. Lyskova\inst{\ref{iki}},
E. Churazov\inst{\ref{mpa}, \ref{iki}},
I. Khabibullin\inst{\ref{rpc}, \ref{mpa}, \ref{iki}},
T. Pasini\inst{\ref{ira}},
E. O'Sullivan\inst{\ref{cfa}},
G. Brunetti\inst{\ref{ira}},
F. De Gasperin\inst{\ref{ira}},
E. De Rubeis\inst{\ref{hamburg}, \ref{ira}},
F. Gastaldello\inst{\ref{iasf}},
D. N. Hoang\inst{\ref{tls}},
R. Kraft\inst{\ref{cfa}},
G. Schellenberger\inst{\ref{cfa}},
R. Sunyaev\inst{\ref{iki}, \ref{mpa}},
R. J. van Weeren\inst{\ref{leiden}},
F. Vazza\inst{\ref{unibo}, \ref{ira}}
}

\institute{
INAF - IRA, via P.~Gobetti 101, 40129 Bologna, Italy \label{ira} \\
\email{andrea.botteon@inaf.it} 
\and
Center for Astrophysics | Harvard \& Smithsonian, 60 Garden Street, Cambridge, MA 02138, USA \label{cfa}
\and
Space Research Institute (IKI), Profsoyuznaya 84/32, Moscow 117997, Russia \label{iki}
\and    
Max Planck Institute for Astrophysics, Karl-Schwarzschild-Str. 1, Garching b. M\"{u}nchen 85741, Germany \label{mpa}
\and
Rudolf Peierls Centre for Theoretical Physics, Department of Physics, University of Oxford, Clarendon Laboratory, Parks Rd, Oxford, OX1 3PU, United Kingdom \label{rpc}
\and
INAF - IASF Milano, via A.~Corti 12, 20133 Milano, Italy \label{iasf}
\and
Hamburger Sternwarte, Universit\"{a}t Hamburg, Gojenbergsweg 112, 21029 Hamburg, Germany \label{hamburg}
\and
Th\"{u}ringer Landessternwarte, Sternwarte 5, 07778 Tautenburg, Germany \label{tls}
\and
Leiden Observatory, Leiden University, PO Box 9513, 2300 RA Leiden, The Netherlands \label{leiden}
\and
Dipartimento di Fisica e Astronomia, Universit\`{a} di Bologna, via P.~Gobetti 93/2, 40129 Bologna, Italy \label{unibo}
}

\date{Received XXX; accepted YYY}

\abstract
{
By inspecting data from the \lotssE\ (\lotss), we noticed a peculiar bright and filamentary radio source at low-galactic latitude ($b \approx 0.5 \deg$). This source, detected also in previous radio observations, was originally believed to be a pulsar, but was later suggested to be located in a heavily obscured galaxy cluster behind the Galactic plane. In this paper, we characterize for the first time the main properties of the host cluster (redshift, mass, temperature, X-ray luminosity, and dynamical status) by using X-ray observations performed with \chandra\ and SRG/\erosita. In addition, by combining new \ugmrt\ follow-up data with observations from the \lolssE\ (\lolss), we perform a multifrequency, spatially resolved spectral analysis of the filamentary radio source (\sou, nicknamed here the ``\treble'' due to its morphology). We conclude that this source is a radio phoenix belonging to a massive, merging galaxy cluster in the Zone of Avoidance. We speculate that its complex morphology is shaped by gas motions generated in the intracluster medium during the ongoing merger, which are also likely responsible for the generation of the candidate radio halo tentatively observed in the cluster center. Owing to its highly filamentary morphology, brightness at $\lesssim$1 GHz, and extremely steep spectrum, reaching values of $\alpha > 4$ between 400 and 650 MHz, this source represents an ideal target for high-resolution, very-low-frequency follow-up observations with \lofar2.0.
}


\maketitle

\section{Introduction}

\begin{figure}
  \centering
  \includegraphics[width=\hsize,trim={0cm 0cm 0cm 0cm},clip,valign=c]{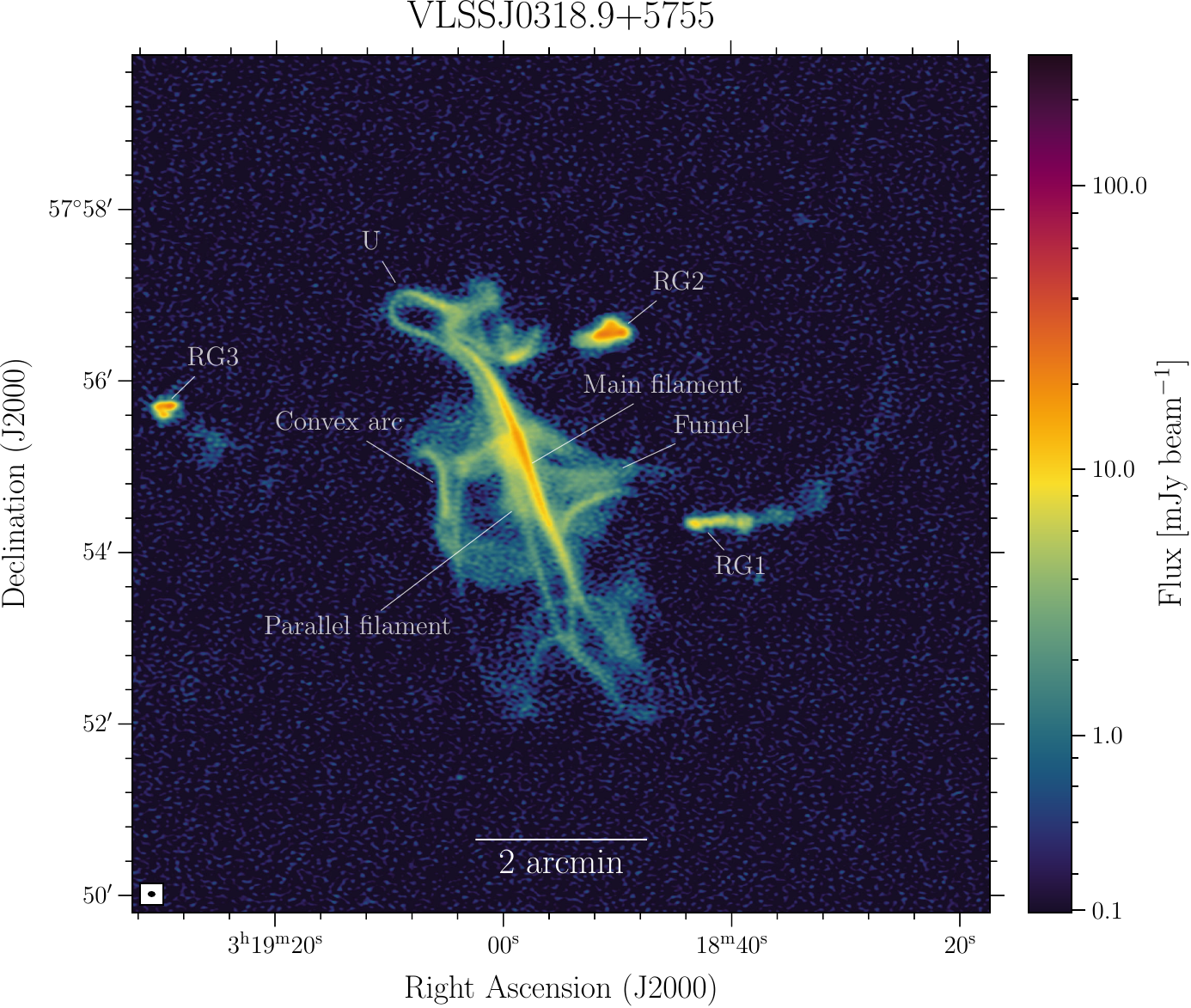}
  \caption{\lofar\ image of the \treble\ (\sou) at 4.1 arcsec $\times$ 2.9 arcsec resolution obtained by recalibrating the data from the \lotss-DR3 \citep{shimwell26}. The main features discussed in the paper are labeled.}
  \label{fig:lofar_highres}
\end{figure}

Diffuse radio emission in galaxy clusters reveals the presence of relativistic cosmic rays (CRs) and magnetic fields permeating the intracluster medium (ICM) \citep[\eg,][for reviews]{brunetti14rev, vanweeren19rev}. These nonthermal components are typically observed as steep-spectrum radio sources (with $\alpha > 1$, where $S_\nu \propto \nu^{-\alpha}$) and are therefore best detected at low frequencies. Thanks to the recent advances in low-frequency radio astronomy, and particularly thanks to the \lofarE\ \citep[\lofar;][]{vanhaarlem13}, we are now able to obtain high-sensitivity and high-resolution images at frequencies $\lesssim$200 MHz \citep{vanweeren16calibration, degasperin20toothbrush, morabito22, sweijen22, dejong22, groeneveld24}. These capabilities have unveiled a rich morphological complexity of radio sources, frequently resolving diffuse emission into filamentary and interconnected substructures that were previously undetected \citep[\eg,][]{hardcastle19, botteon20a2255, brienza21, rajpurohit22a2256, mahatma23, derubeis25a2255}. \\
\indent
A particularly intriguing class of sources that exhibits both ultra-steep spectra and filamentary morphology and whose population is rapidly growing especially thanks to new low-frequency observations is that of radio phoenixes \citep[\eg,][]{kempner04taxonomy, vanweeren11vlss, kale18, mandal20, bruno25}. These are diffuse, relatively compact (few hundred kpc) sources, characterized by ultra-steep and often curved spectra (typical spectral indexes of $\alpha > 1.5$). In the currently favored scenario, radio phoenixes trace fossil plasma \citep[also referred to as ``remnant'' plasma, \eg,][]{murgia11, brienza17, quici21} produced by earlier episodes of active galactic nucleus (AGN) jet activity. This plasma has faded due to radiative losses and is later compressed and/or reenergized by dynamical processes in the cluster (such as mergers, shocks, or gas motions), making it visible again in the radio band and giving rise to sources with complex morphologies \citep[\eg,][]{ensslin01, ensslin02relics}. An archetypal example is the phoenix in the galaxy cluster Abell 85 \citep{slee01}, whose recent observations have revealed its intricate filamentary structure in exceptional detail \citep{raja24a85}, demonstrating the step forward in imaging fidelity provided by modern radio instruments. \\
\indent
Filamentary structures in cluster radio sources offer a unique laboratory for studying the propagation and confinement of CRs within the ICM. The existence of thin, magnetized filaments implies that CR transport can be highly constrained, with particles likely confined to magnetic flux tubes rather than freely diffusing through the turbulent and high-$\beta$ plasma\footnote{Defined as the ratio between the thermal and magnetic pressures, that is $\beta_{\rm pl} = P_{\rm th}/P_{\rm B}$.} ($\beta_{\rm pl} \sim 10^2$) of the ICM. These structures may trace localized sites of particle reacceleration and/or compression \citep{rudnick22, alam25filament, brienza25, derubeis26}, potentially associated with turbulence, weak shocks, or magnetic reconnection. Another possibility is that transient electric fields induced by variable relativistic jets could trigger localized discharges along the filaments, illuminating them with freshly accelerated electrons \citep{gopalkrishna24threads}. However, it has also been proposed that in nonthermal filaments characterized by $\beta_{\rm pl} \lesssim 1$, CR propagation can be fast and effectively ``lossless'', occurring at super-Alfvénic speeds in terms of the Alfvén speed in the thermal ICM \citep{churazov26filaments}. In this scenario, additional reacceleration mechanisms are not required to explain the large extent (often $\sim$10$^2$ kpc) of these thin structures. Detailed studies of radio filaments therefore can provide important constraints on CR transport and reacceleration processes in the ICM, allowing us to investigate the interplay between thermal and nonthermal components in galaxy clusters. \\
\indent
While inspecting observations of the \lotssE\ \citep[\lotss;][]{shimwell19, shimwell22}, in one of the Galactic fields recently released with the third data release \citep[\lotss-DR3;][]{shimwell26}, we noticed a bright and extended ($\approx$6 arcmin along its major axis) radio source with a peculiar morphology, shown in Fig.~\ref{fig:lofar_highres}. A cross-match with existing radio catalogs revealed that this emission is listed in the \vlssE\ \citep[\vlss;][]{cohen07} with the name \sou. A literature search led us to the paper by \citet{green04}, the most recent work in which this source\footnote{In \citet{green04}, the source is reported with the name WKB 0314+57.8, following its original identification by \citet{williams66}.} is discussed in detail; we refer the reader to that paper for a comprehensive overview of its previous detections in other radio surveys. For the purposes of this work, the key results from the broadband radio and near-infrared (NIR) observations presented by \citet{green04} are that the source is resolved (although its detailed morphology could not be established because of the limited angular resolution available at the time), has an integrated ultra-steep spectrum of $\alpha = 2.5$ between 40 MHz and 1.5 GHz, and is tentatively associated with a galaxy cluster located behind the Galactic plane\footnote{Owing to its ultra-steep spectrum and low Galactic latitude ($b \approx 0.5 \deg$), the source was initially classified as a pulsar candidate \citep{erickson85}.} likely at a redshift of $z \approx 0.08$, as estimated from the NIR color-magnitude diagram analysis. Since then, the source was not further studied, with no dedicated follow-up observations reported. To the best of our knowledge, it has only been mentioned once more in the literature, in the catalog of candidate X-shaped radio sources compiled by \citet{bhukta22xzshaped} using the \tgssE\ Alternative DR1 \citep[\tgss-ADR1;][]{intema17}. \\
\indent
Our \lofar\ image shown in Fig.~\ref{fig:lofar_highres} provides an unprecedented view of \sou, revealing a complex morphology characterized by numerous filaments and loop-like structures that led us to nickname it the ``\treble''. The intricate structure of the emission, together with its ultra-steep spectrum, would be consistent with its classification as a radio phoenix. Indeed, the \treble\ shows some morphological similarities with the already cited archetypal radio phoenix in Abell~85 \citep{raja24a85}. To further investigate the nature and origin of this source, we obtained follow-up observations with the \ugmrtE\ (\ugmrt) in band 3 and band 4, and we also made use of data from the forthcoming second data release of the \lolssE\ (\lolss-DR2; De Gasperin et al., in prep.). In this paper, we present the analysis of these radio observations, which we complement with X-ray data from \chandra\ and \erositaE\ (\erosita) to characterize the main properties of the host cluster for the first time. \\
\indent
Hereafter, we adopt a $\Lambda$ cold dark matter cosmology with $\omegal = 0.7$, $\omegam = 0.3$, and $\hzero = 70$ \kmsmpc.

\begin{table*}[t]
\centering
\caption{Overview of the radio observations.}
\label{tab:radio_overview}
\resizebox{\hsize}{!}{
\begin{tabular}{lccccccc}
\hline
\hline
  Instrument & Project & Pointing & Observing date & Frequency range & Channel width & Time on source\\
   &  &  & (dd-mm-yyyy) & (MHz) & (kHz) & (h) \\
\hline
  \lofar\ LBA & LT14\_002 & P049+59 & 26-08-2021, 06/12-08-2022, 22-04-2023 & 42--66 & 3.05  & 4 \\
  \lofar\ LBA & LT14\_002 & P051+56 & 09/29-03-2022, 20-04-2022 & 42--66 & 3.05  & 3 \\
  \lofar\ HBA & LC12\_014 & P049+59 & 28-11-2019 & 120--168 & 12.19 & 8 \\
  \lofar\ HBA & LT14\_004 & P051+56 & 13-02-2022 & 120--168 & 12.19 & 8 \\
  \lofar\ HBA & LT14\_004 & P046+56 & 13-12-2021 & 120--168 & 12.19 & 8 \\
  \ugmrt\ band 3 & 44\_033 & -- & 03-09-2023 & 300--500 & 48.83 & 6 \\
  \ugmrt\ band 4 & 44\_033 & -- & 05-09-2023 & 550--950 & 97.66 & 6 \\
\hline
\end{tabular}
}
\tablefoot{Offsets of the \treble\ from the LOFAR pointing centers are: 1.1\deg\ (P049+59), 1.7\deg\ (P051+56), and 2.2\deg\ (P046+56).}
\end{table*}

\begin{table}[t]
 \centering
 \caption{Properties of the radio images.}
 \label{tab:radio_images}
 \resizebox{\hsize}{!}{
  \begin{tabular}{lccccc}
  \hline
  \hline
  Instrument & Frequency & Taper & Beam & rms & Figure \\
   & (MHz) &  (\arcsec) & ($\arcsec \times \arcsec$)& (\mujyb) & \\
  \hline
  \lofar\ LBA & 54 & -- & $10 \times 10$ & 5098 & \ref{fig:4panels} \\
  \lofar\ HBA & 145 & -- & $4.1 \times 2.9$ & 191 & \ref{fig:lofar_highres}, \ref{fig:overlays}, \ref{fig:2panels} \\
  \lofar\ HBA & 145 & 8 & $10 \times 10$ & 188 & \ref{fig:4panels} \\
  \ugmrt\ band 3 & 400 & 8  & $10 \times 10$ & 24 & \ref{fig:4panels} \\
  \ugmrt\ band 4 & 650 & 8  & $10 \times 10$ & 18 & \ref{fig:4panels} \\
  \hline
  \end{tabular}
  }
  \tablefoot{All images were obtained with a robust weighting of $-1.5$ \citep{briggs95}. Circular beams were obtained by applying a small amount of Gaussian smoothing to the images to reach the desired resolution.}
\end{table}

\section{Data reduction}

In the following sections, we describe the data reduction procedures for the observations used in this work. An overview of the radio observations is provided in Tab.~\ref{tab:radio_overview}. For all calibrated datasets, final imaging was performed with \wsclean\ v3.6 \citep{offringa14} using multiscale, multifrequency deconvolution \citep{offringa17}. Images used for spectral analysis were aligned, regridded, and smoothed to a common angular resolution. Details of the radio images presented in this paper are summarized in Tab.~\ref{tab:radio_images}.

\subsection{\lofar\ LBA and HBA}

We made use of \lofar\ Low/High Band Antenna (LBA/HBA) observations from the \lolss\ \citep{degasperin23} and \lotss\ \citep{shimwell19} low-frequency surveys of the northern sky. For the LBA data, covering the frequency range 42--66 MHz, we used two \lolss\ pointings that include the \treble\ and that will be publicly released with \lolss-DR2 (De Gasperin et al., in prep.). For the HBA data, covering 120--168 MHz, we used three pointings from the recently released \lotss-DR3 \citep{shimwell26}. The data were calibrated by the \lofar\ Surveys Key Science Project team using dedicated pipelines developed for the two surveys \citep{vanweeren16calibration, williams16, degasperin19, degasperin21, tasse21}. \\
\indent
In this work, we adopt the now standardized ``extraction+recalibration'' reprocessing scheme of \lofar\ observations introduced in \citet{vanweeren21}. Briefly, the extraction step consists of removing from the \uv\ plane sources outside a region of interest containing the target, phase-shifting the visibilities to the center of this region, and correcting for the \lofar\ primary beam in that direction. After that, the data are averaged in time and frequency. During the subsequent recalibration step, the data undergo multiple self-calibration loops to optimize the calibration quality of the target source. This approach corrects residual direction-dependent errors from the survey processing pipelines and enables flexible and efficient imaging of the smaller, recalibrated dataset. \\
\indent
We adopted a systematic uncertainty on the LBA (54 MHz) and HBA (145 MHz) flux densities of 6\% \citep{degasperin23} and 10\% \citep{shimwell26}, respectively.

\subsection{\ugmrt\ band 3 and 4}

We followed-up the \treble\ with targeted \ugmrt\ observations in band 3 and 4 during Cycle 44. Data were recorded using both the GMRT Software Backend \citep[GSB;][decommissioned from Cycle 47]{roy10} and GMRT Wideband Backend \citep[GWB;][]{reddy17} correlators. Data were processed using the \spamE\ \citep[\spam;][]{intema09} pipeline, which performs a fully automated end-to-end calibration, including bandpass calibration, flagging, averaging, and both direction-independent and direction-dependent self-calibration. The narrowband GSB data were only used to obtain an initial sky model for the wideband GWB observations, which are used to perform the analysis presented in the paper. The \spam\ pipeline handles the GWB data by splitting the wideband observations into narrow frequency slices which are calibrated independently and then jointly deconvolved with \wsclean\ to produce the final images. For the band~4 observations, we excluded data at frequencies $>$750~MHz from the analysis due to its significantly lower quality. \\
\indent
We adopted a systematic uncertainty on the band 3 (400 MHz) and band 4 (650 MHz) flux densities of 10\% and 5\% \citep{chandra04}, respectively.

\subsection{Chandra}

We used the \chandra\ Guaranteed Time Observer program to observe the target with the \acisE\ (\acis) in April 2026. The observations consist of four runs (ObsID: 30487, 32286, 32287, 32288), totaling 48.8~ks of on-source time, which were processed with the standard \chandra\ data reduction software, using \ciao\ v4.17 and CALDB v4.12 \citep{fruscione26ciao}. Data analysis steps are described in detail in \citet{vikhlinin09ii}  and include high background period filtering, application of the latest calibration corrections to the detected X-ray photons, and determination of the background intensity in each observation. Additionally, we filtered out time intervals when the \acis\ focal plane temperature was above 162 K (see Appendix~\ref{app:ftemp} for the impact of this on spectral results). For spectral analysis, we generated the spectral response files that combine the position-dependent \acis\ calibration with the weights proportional to the observed brightness. The total filtered exposure time is $\approx$25.9 ks.

\subsection{eROSITA}

We use the data from the \erosita\ telescope \citep{predehl21} on board the \srgE\ (SRG) mission \citep{sunyaev21}, launched in 2019, which started to perform the all-sky survey mission in December 2019. We use the data accumulated over four consecutive scans, with the total effective exposure amounting to $\approx$1000~s per point. Initial reduction and processing of the data were performed using standard routines of the eROSITA Science Analysis Software System \citep[eSASS;][]{brunner18, predehl21}, while the imaging and spectral analysis were carried out with the background models, vignetting, point spread function and spectral response function calibrations built upon the standard ones via slight modifications motivated by results of calibration and performance verification observations \citep[\eg,][]{churazov21, khabibullin23}.

\section{Is there a galaxy cluster at the location of the \treble?}\label{sec:xray}

\begin{figure}
  \centering
  \includegraphics[width=\hsize,trim={0cm 0cm 0cm 0cm},clip,valign=c]{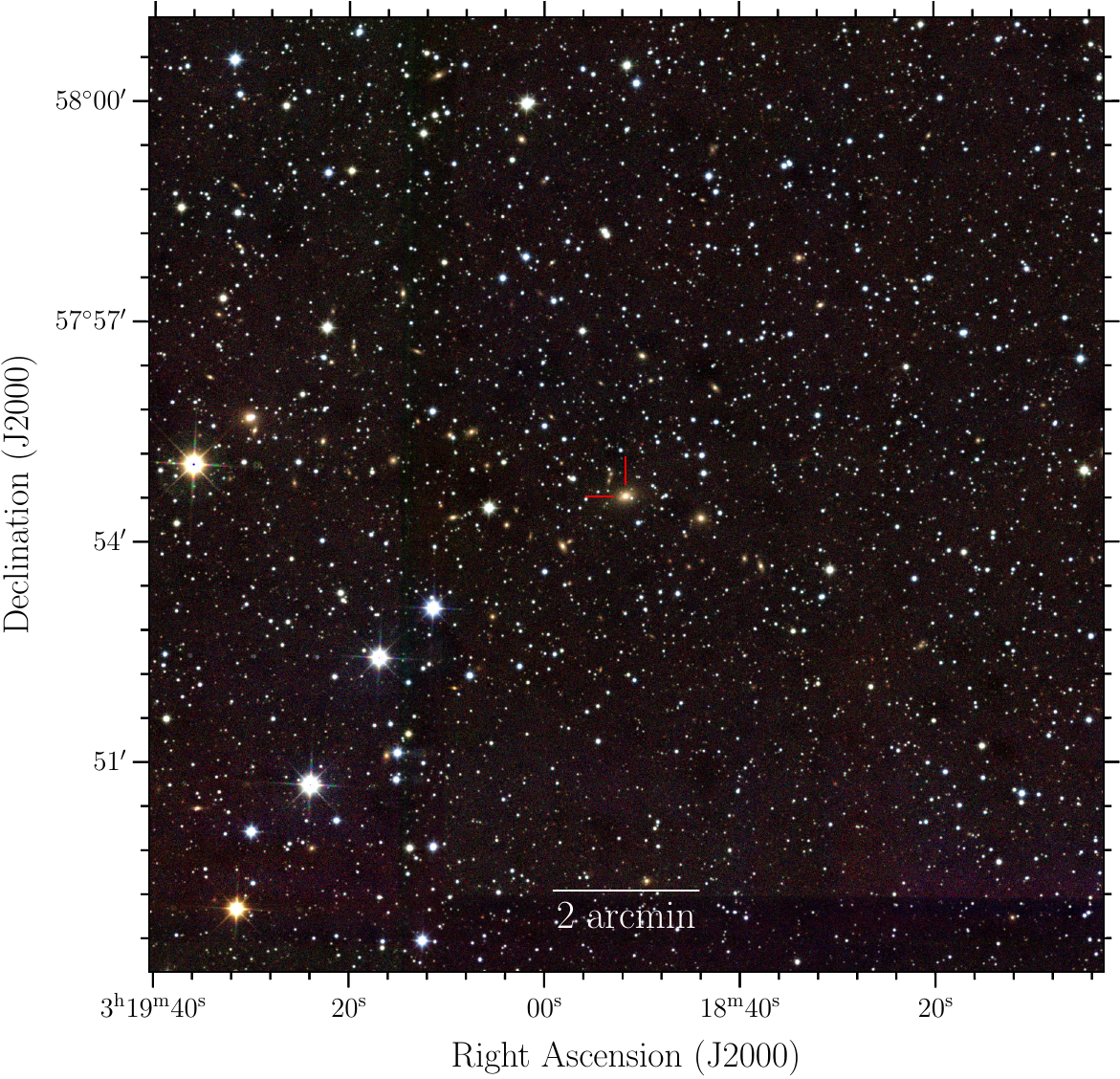}
  \caption{\ukidss-GPS \textit{JHK} composite image of the field around the \treble. The FoV is 13 arcmin $\times$ 13 arcmin and the position of the candidate BCG (WISEA J031851.64+575437.7) is marked by two red segments.}
  \label{fig:ukidss}
\end{figure}

Indications of a possible galaxy cluster at the location of the \treble\ were proposed by \citet{green04} from the NIR color-magnitude diagram obtained with the Lick Observatory. In Fig.~\ref{fig:ukidss}, we report a NIR \textit{JHK} image obtained from the \ukidssE-Galactic Plane Survey \citep[\ukidss-GPS;][]{lawrence07, lucas08} of a field-of-view (FoV) centered on the brightest galaxy nearby the \treble. This galaxy, marked by two segments, is located at RA = 49.715\deg\ and Dec = 57.911\deg\ and is reported with the name WISEA J031851.64+575437.7 in the AllWISE Source Catalog \citep{wright10, mainzer11}. Despite the crowded field due to the low Galactic latitude, several galaxies with similar colors can be visually identified in its vicinity, suggesting the presence of an overdensity and that this galaxy likely corresponds to the brightest cluster galaxy (BCG). 

\begin{figure}
  \centering
  \includegraphics[width=\hsize,trim={0cm 0cm 0cm 0cm},clip,valign=c]{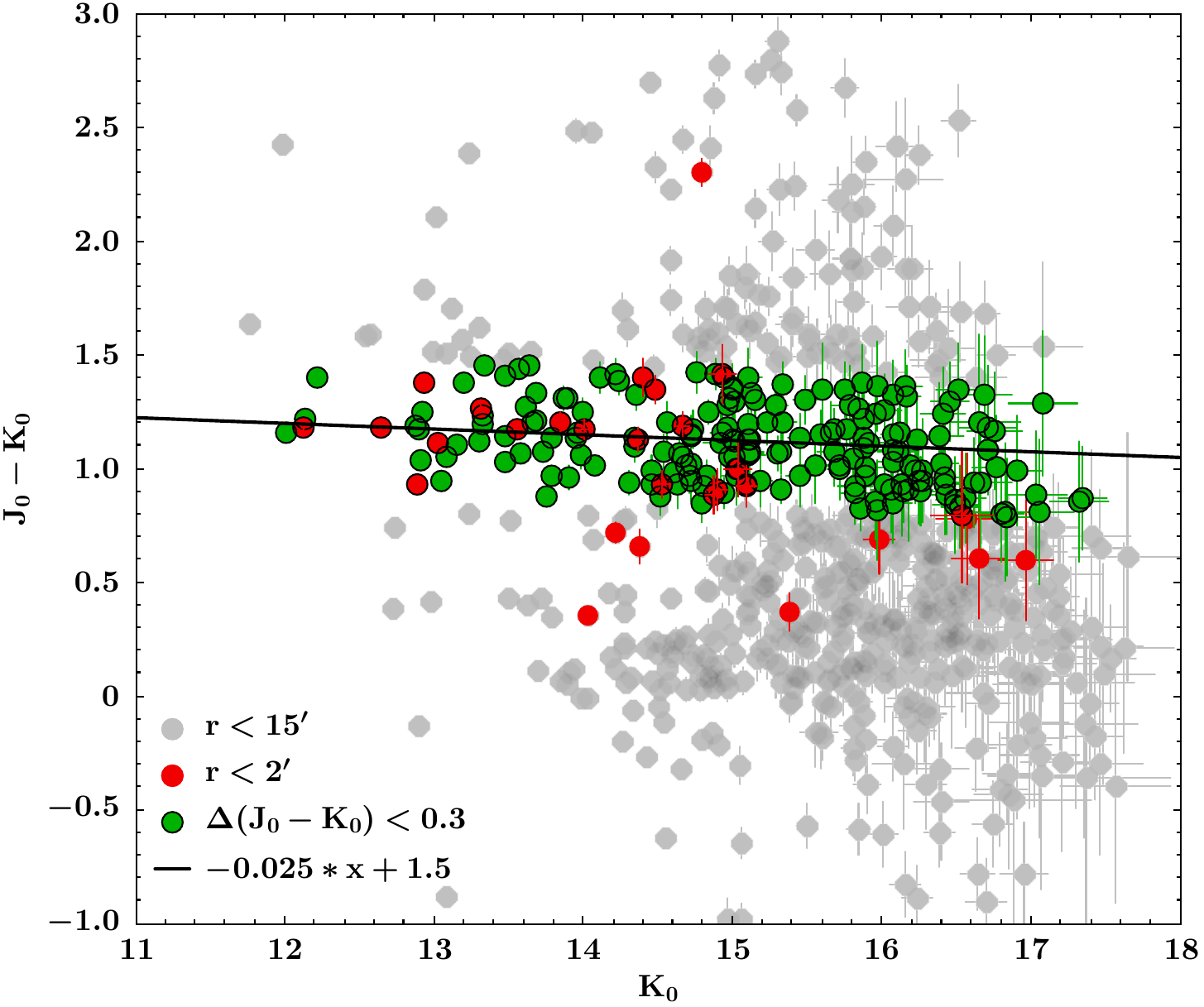}
  \caption{Color-magnitude diagram obtained from the \ukidss-GPS DR11 catalog \citep{lucas08} for galaxies within 2 arcmin (red) and 15 arcmin (gray) of the center of X-ray emission. The black line shows the absorption-corrected NIR red sequence and green circles denote galaxies within a color-difference of $\Delta(J_0 - K_0 ) < 0.3$ from it.}
  \label{fig:color-mag}
\end{figure}

\begin{figure*}
 \centering
 \includegraphics[width=0.49\hsize]{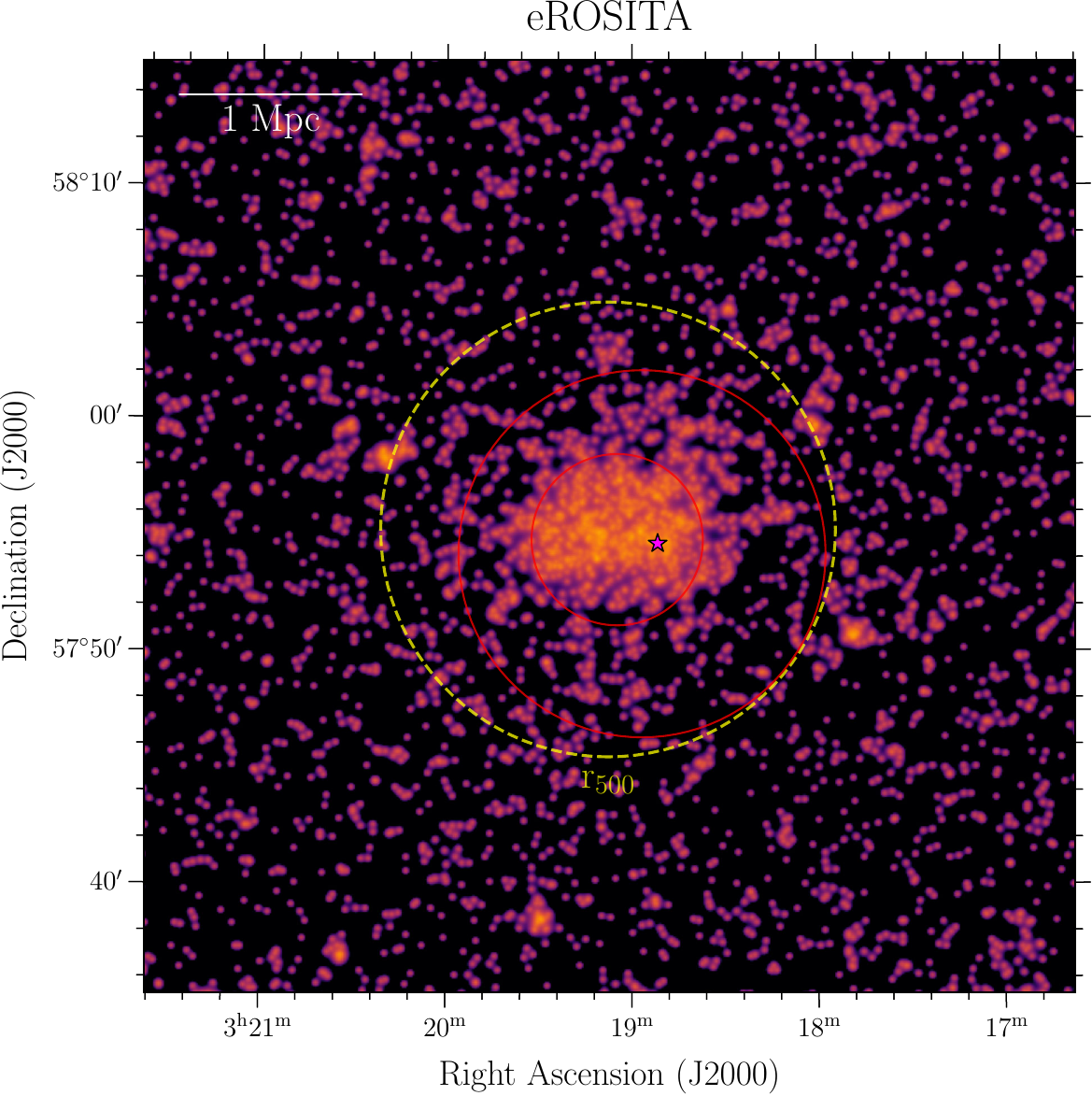}
 \includegraphics[width=0.49\hsize]{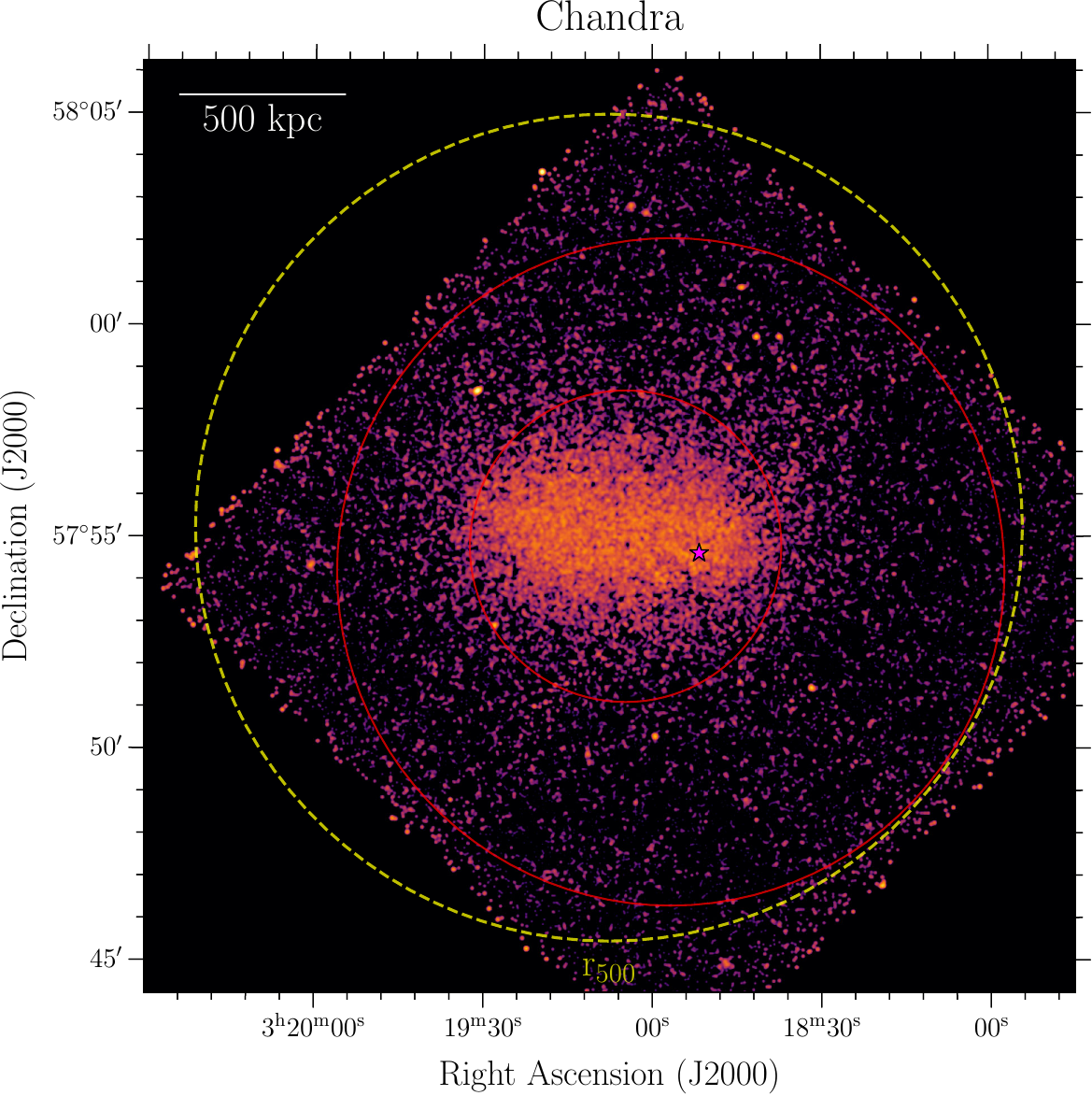}
 \caption{\erosita\ 0.3--2.3 keV (\textit{left}) and \chandra\ 1--5 keV (\textit{right}) images of the cluster. The yellow dashed circle corresponds to \rfive\ if \mfive\ is calculated from the $\mfive$--$T$ relation of \citet{vikhlinin09ii}. Spectra shown in Fig.~\ref{fig:xspec} are extracted from the inner circle of radius 3.7 arcmin using the region between two red circles as background. The magenta star marks the position of the BCG. Images are centered at the position adopted as center for the surface brightness profile modeling.}
 \label{fig:ximage}
\end{figure*}

\begin{figure}
 \centering
 \includegraphics[width=\hsize]{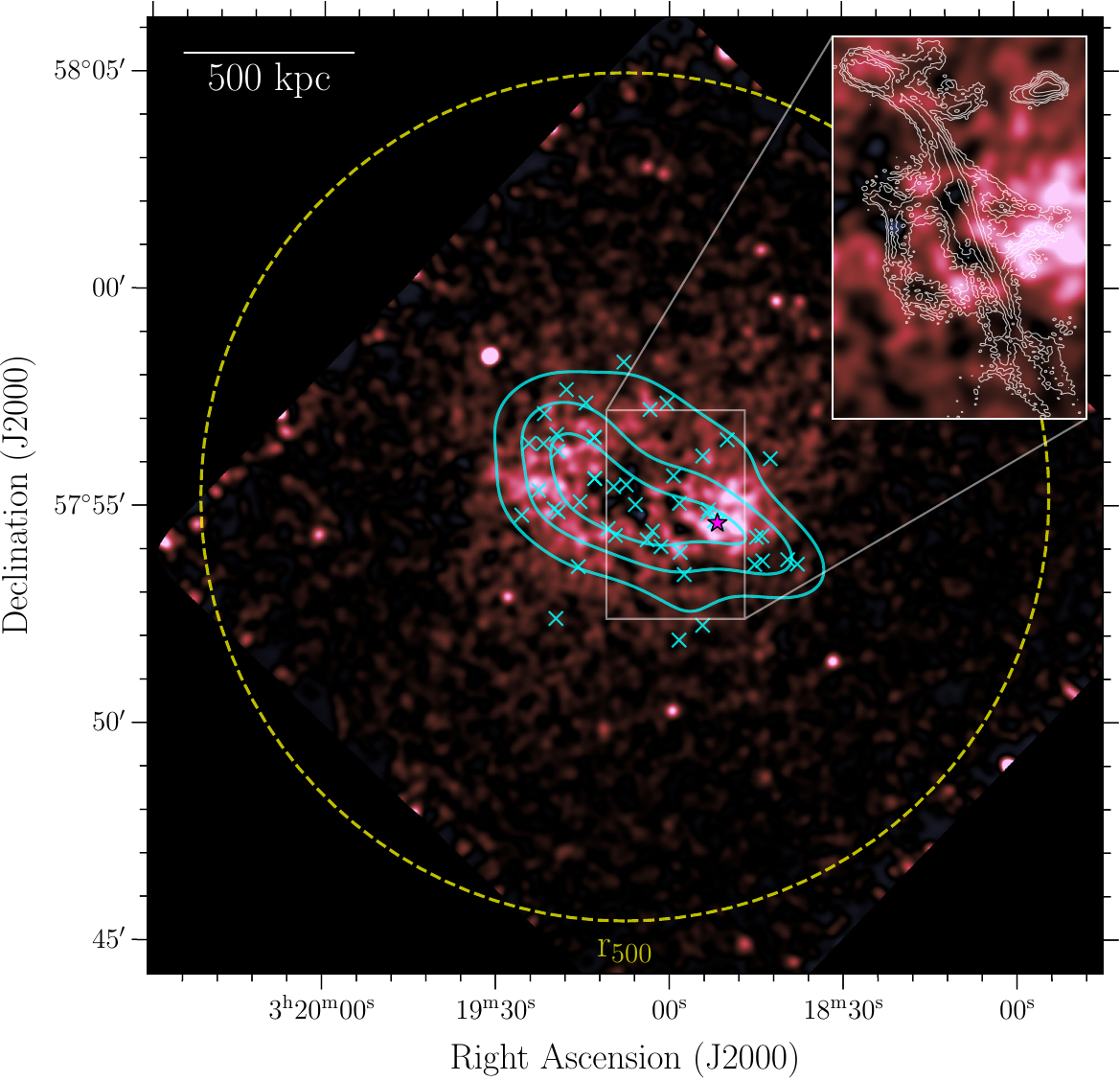}
 \caption{Residuals of the best-fit $\beta$-model on the \chandra\ data. The cyan contours are the 2D kernel density estimate of the galaxies (denoted with crosses) with a color difference of $<$0.3 from the red sequence that are located within 500 kpc from the cluster center. The magenta star marks the position of the BCG. The inset shows the \lofar\ contours overlaid in the region with indication of thermal gas depletion.}
 \label{fig:residuals}
\end{figure}

\begin{figure}
 \centering
 \includegraphics[width=\hsize]{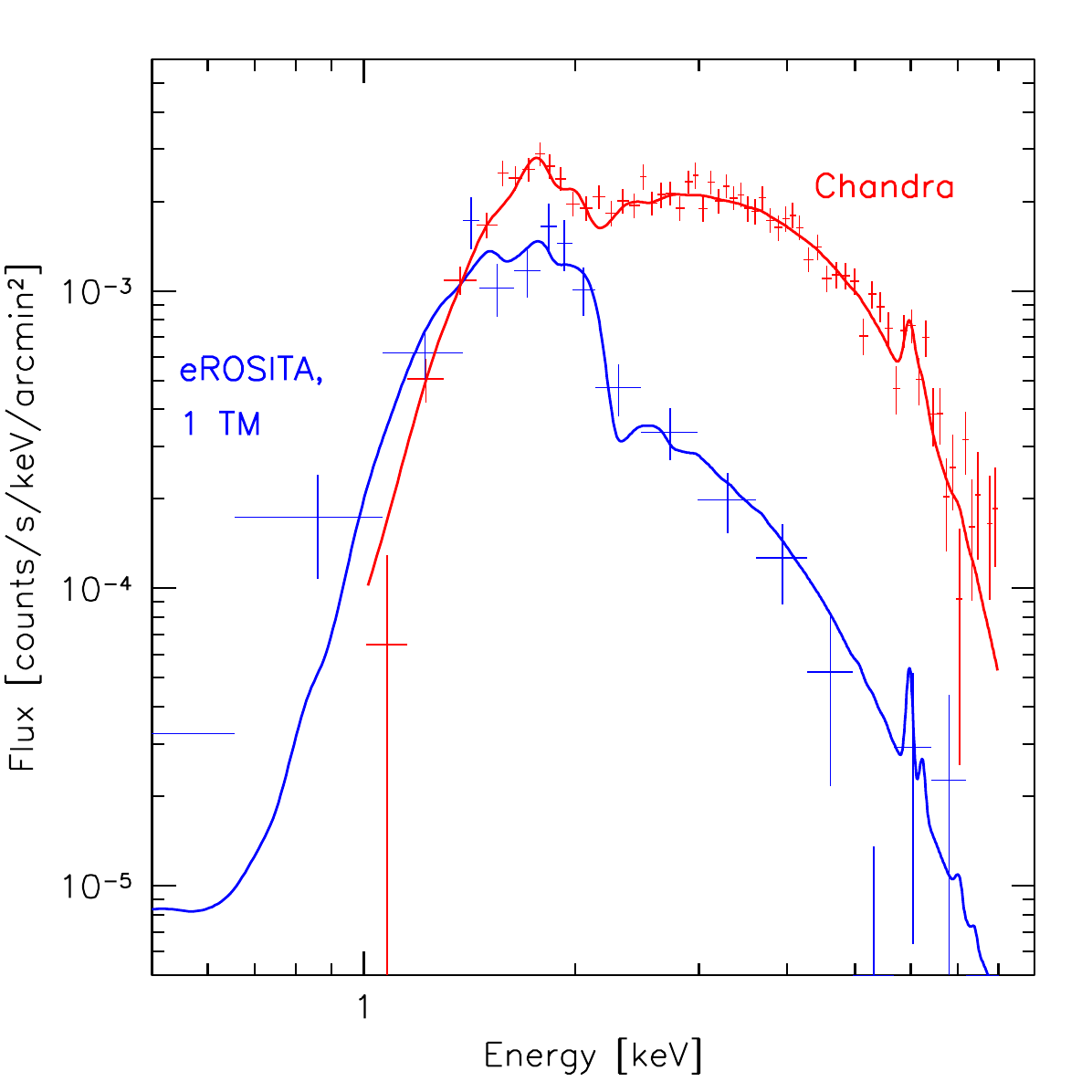}
 \caption{\chandra\ and \erosita\ spectra of the cluster. The \erosita\ spectrum is normalized per one (out of seven) \erosita\ telescope modules. The best-fit absorbed APEC model obtained from the joint analysis of \chandra +\erosita\ data is shown with solid curves. The red and blue curves correspond to the same spectral model convolved with the \chandra\ and \erosita\ responses, respectively.}
 \label{fig:xspec}
\end{figure}

\begin{table*}
 \centering
 \caption{Properties of the cluster derived from X-ray observations.}
 \label{tab:cluster_properties}
  \begin{tabular}{cccccccc}
  \hline
  \hline
  $z$ & $kT$ & $Z$ & $L_{\rm [0.5-2.0\:keV]}$ & $M_{500}^{\rm L-based}$ & $M_{500}^{\rm T-based}$ & $c$ & $w$ \\
      & (keV) & ($Z_\odot$) & (\ergs) & (\msun) & (\msun) & ($\times 10^{-2}$) & ($\times 10^{-3}$) \\
  \hline
  $0.117\pm0.008$ & $6.4\pm0.6$ & $0.22\pm0.05$ & $2.2\times 10^{44}$ & $5.4\times 10^{14}$ & $6.0\times 10^{14}$ & $8.84\pm0.27$ & $8.01\pm1.08$ \\
  \hline
  \end{tabular}
  \tablefoot{Reported quantities are redshift ($z$), temperature ($kT$), metallicity ($Z$), X-ray luminosity in the 0.5--2.0 \kev\ band ($L_{\rm [0.5-2.0\:keV]}$), mass derived from the mass--luminosity ($M_{500}^{\rm L-based}$) and mass--temperature ($M_{500}^{\rm T-based}$) relations, concentration parameter ($c$), and centroid shift ($w$).}
\end{table*}

\begin{figure*}
  \centering
  \includegraphics[width=.49\hsize,trim={0cm 0cm 0cm 0cm},clip,valign=c]{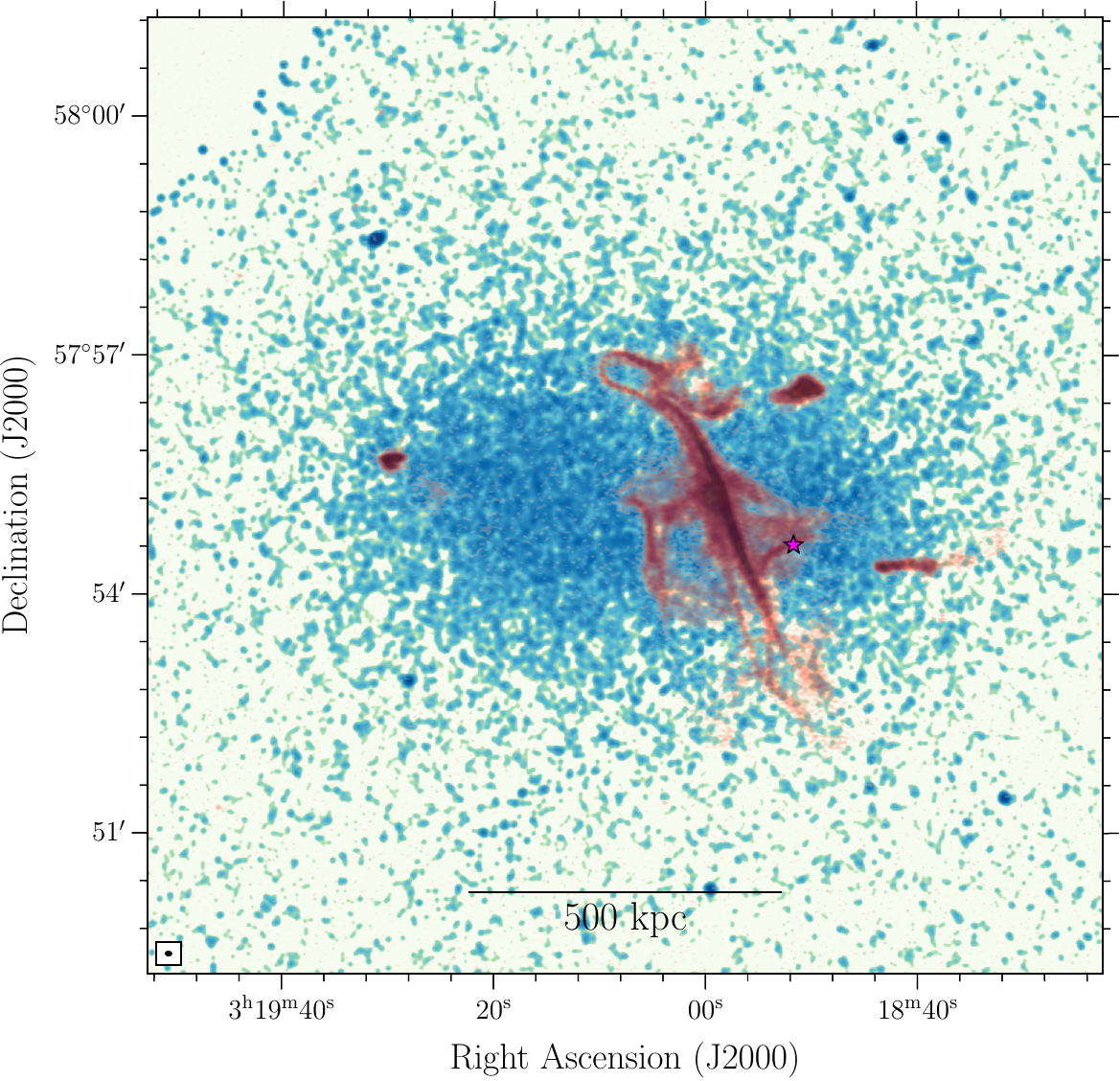}
  \includegraphics[width=.49\hsize,trim={0cm 0cm 0cm 0cm},clip,valign=c]{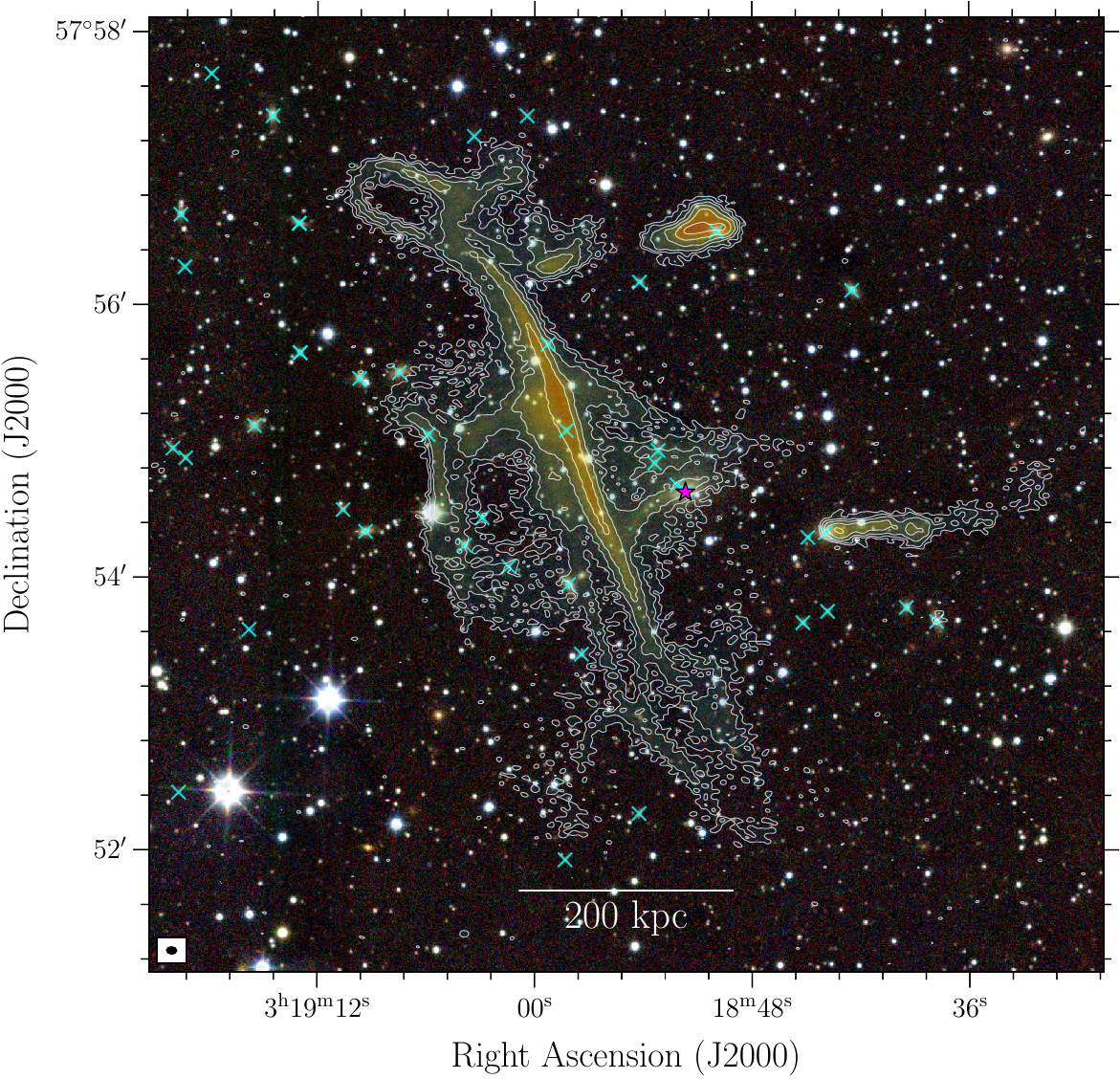}
  \caption{\chandra-\lofar\ (\textit{left}) and \ukidss-\lofar\ (\textit{right}) overlays. The displayed \lofar\ image is the same as in Fig.~\ref{fig:lofar_highres}. Cyan crosses are the same of Fig.~\ref{fig:residuals}, with the magenta star marking the position of the BCG (in the left panel, only this one is reported).}
  \label{fig:overlays}
\end{figure*}

Similarly to previous studies of galaxy clusters in the Zone of Avoidance \citep[\eg,][]{ramatsoku20nir, khabibullin26arx}, we constructed the (absorption-corrected) NIR color-magnitude diagram $(J_0-K_0)$--$K_0$ shown in Fig.~\ref{fig:color-mag} using all bona-fide extended sources in the \ukidss-GPS DR11 catalogue. A subsample of galaxies located within 2 arcmin from the center was used to identify the cluster red sequence, which turned out to be consistent with the relation $(J_0-K_0)=-0.025K_0+1.5$, virtually identical to the one found for other clusters \citep[\eg,][]{ramatsoku20nir, khabibullin26arx}. Galaxies within much larger ($r<15$ arcmin) region with a color difference smaller than 0.3 with respect to this relation are highlighted in green in this plot and can be considered as candidate member galaxies in future studies. \\
\indent
The presence of extended X-ray emitting gas detected with \chandra\ and eROSITA in Fig.~\ref{fig:ximage} unambiguously confirms the presence of a galaxy cluster in the field. The emission from the newly detected system appears slightly elongated in the east-west direction ($\approx$10 arcmin $\times$ 8 arcmin), its center is offset with respect to the position of the BCG, and it lacks a pronounced central surface brightness peak. The X-ray morphological parameters $c$ (concentration, \eg, \citealt{santos08}) and $w$ (centroid shift, \eg, \citealt{poole06}) computed from the \chandra\ observation following \citet{zhang23} are $c=(8.84\pm0.27) \times 10^{-2}$ and $w=(8.01\pm1.08) \times 10^{-3}$, respectively. These characteristics indicate that the system is dynamically unrelaxed. \\
\indent
We provide a first characterization of the X-ray surface brightness profile by fitting the simplest spherically symmetric $\beta$-model \citep{cavaliere76} to the \chandra\ and \erosita\ data, taking RA $= 49.782\deg$ and Dec $= 57.920\deg$ as the cluster center (since there is no pronounced peak in the X-ray image, the cluster center was defined as an approximate centroid of the X-ray surface brightness distribution on a scale of several arcminutes). The resulting best-fit core radius is $r_{\rm c} = 3.76$ arcmin for \chandra\ (1--5 keV) and $4.98$ arcmin for \erosita\ (0.3--2.3 keV), while the corresponding slope parameters are $\beta = 1.06$ and $\beta = 1.19$, respectively (see Appendix~\ref{app:beta} for details).  Despite the differences in the best-fit $\beta$-model parameters, the inferred central electron density is consistent between the two datasets, with values of $(2\text{--}3) \times 10^{-3}$ cm$^{-3}$ (this estimate uses the redshift derived from the X-ray spectral fitting described below). The residuals of the \chandra\ model shown in Fig.~\ref{fig:residuals} highlight two surface brightness excesses toward the east and west, aligned with the contours tracing the distribution of red-sequence galaxies. These possibly indicate a merger of two main systems occurring primarily along the east--west direction. A hint of X-ray surface brightness decrements cospatial with the brightest region of the \treble\ suggests that the nonthermal plasma may have displaced the thermal gas in a manner similar to AGN-inflated bubbles observed in X-ray cavities (see inset panel). \\
\indent
In order to determine the properties of the cluster, we analyze \chandra\ and \erosita\ spectra extracted from the red circular region of radius 3.7 arcmin showed in Fig.~\ref{fig:ximage}, using the area of the outer annulus as background region\footnote{As will become evident from the analysis presented below, the \chandra\ observation covers only the central part of the cluster, making it impossible to define a background region free from cluster emission. We therefore experimented with several choices of background regions, both in the analysis of the \chandra\ data alone and in the joint analysis of the \chandra\ and eROSITA data. In all cases, the derived spectral parameters remained consistent within their uncertainties.}. The resulting \erosita\ and \chandra\ spectra, in which point sources were subtracted, are shown in Fig.~\ref{fig:xspec}. We fit the spectra in \xspec\ \citep{arnaud96} assuming an absorbed APEC \citep{smith01} model (\verb #phabs*apec#) leaving all the parameters free to vary during the fit. The best-fit model parameters are: $\nh = (1.62 \pm 0.08) \times 10^{22}$ cm$^{-2}$ for the hydrogen column density, $kT = 6.4 \pm 0.6$ keV for the ICM temperature, $Z/Z_{\odot} = 0.22 \pm 0.05$ for the metal abundance (using the abundance table of \citealt{anders89}), and $z = 0.117 \pm 0.008$ for the cluster redshift. The provided uncertainties are of a statistical nature. Given the uncertainties associated with the method used by \citet{green04} to derive the cluster redshift ($z \approx 0.08$), we consider our result broadly consistent with their estimate and adopt $z = 0.117$ in the subsequent analysis. With the adopted cosmology, 1 arcsec corresponds to 2.12 kpc at the cluster redshift. \\
\indent
To calculate the cluster X-ray luminosity in the rest frame, we first derive the luminosity estimate within the region used for the X-ray spectral analysis (inner red circle in Fig.~\ref{fig:ximage}) from the best-fitting spectral model. We then extrapolate this value to the total cluster luminosity assuming the best-fit $\beta$-model ($\beta = 1.06$, $r_{\rm c} = 3.76$ arcmin). The resulting rest-frame cluster (unabsorbed) luminosity in the 0.5--2.0 keV band is $L_{\rm X} \approx 2.2 \times 10^{44}$ \ergs. Note that the complexity of background and foreground near the Galactic plane and the ongoing merger introduce additional uncertainties to the luminosity estimate. We estimate that using the outer annulus shown in Fig.~\ref{fig:ximage} as the background region introduces an uncertainty of $\approx$10\% to the inferred $L_{\rm X}$. Adopting an elliptical instead of a circular aperture to estimate the central luminosity, followed by extrapolation to the total cluster luminosity, changes the inferred $L_{\rm X}$ by a similar amount ($\approx$10\%).\\
\indent
Using the scaling relations from \citet{vikhlinin09ii} for the measured temperature and X-ray luminosity, we estimated the total cluster mass within the radius enclosing a mean mass density equal to 500 times the critical density at the cluster redshift. We obtained $M_{500}^{\rm L-based} \approx 5.4 \times 10^{14}$ \msun\ and $M_{500}^{\rm T-based} \approx 6.0 \times 10^{14}$ \msun\ from the luminosity- and temperature-based relations, respectively. \\
\indent
These results suggest that the \treble\ belongs to an overlooked massive, merging galaxy cluster located in the Zone of Avoidance, whose main properties derived in this section are summarized in Tab.~\ref{tab:cluster_properties}.

\section{Radio properties}

\begin{figure*}
  \centering
  \includegraphics[width=\hsize,trim={0cm 0cm 0cm 0cm},clip,valign=c]{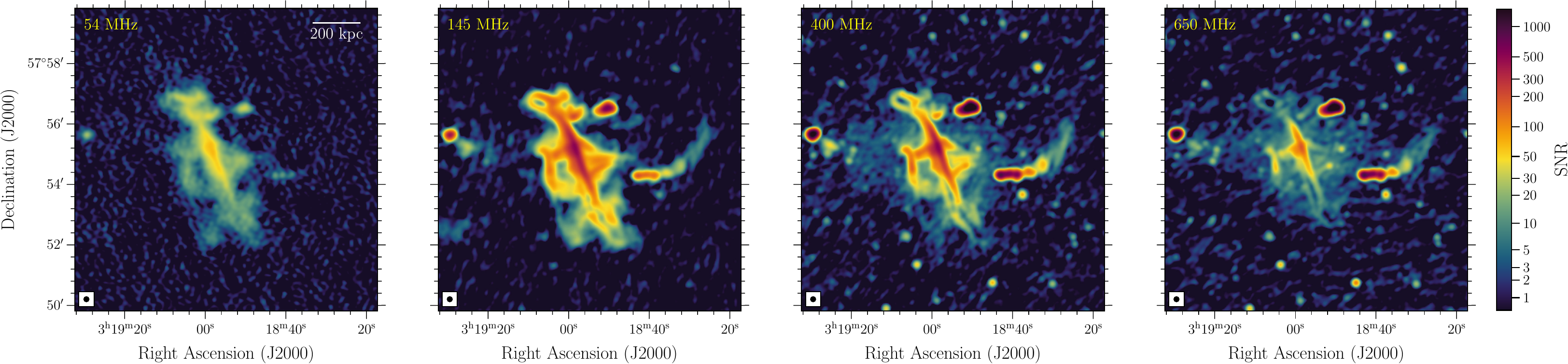}
  \caption{Multi-frequency radio observations. Images have a common resolution of 10 arcsec and show the radio emission as detected with \lofar\ at 54 and 145 MHz and with the \ugmrt\ at 400 and 650 MHz (from \textit{left} to \textit{right}). The color scale has a logarithmic stretch from 0.5 to 1500$\sigma$ of each individual image in all panels (see Tab.~\ref{tab:radio_images} for the rms values).}
  \label{fig:4panels}
\end{figure*}

\begin{figure*}
  \centering
  \includegraphics[width=\hsize,trim={0cm 0cm 0cm 0cm},clip,valign=c]{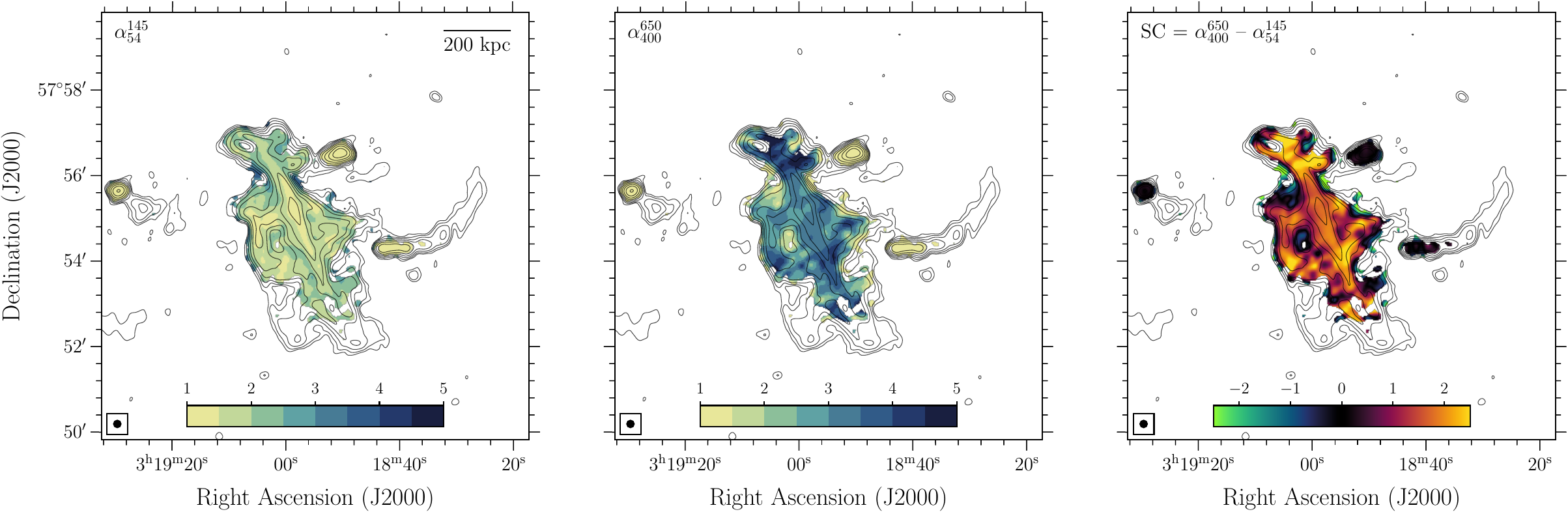}
  \caption{Resolved radio spectral analysis at 10 arcsec resolution. Images show the spectral index maps at low (54--145 MHz, \textit{left}) and high (400--650 MHz, \textit{center}) frequency, and the corresponding spectral curvature map (\textit{right}). The corresponding error maps are reported in Fig.~\ref{fig:3panels_alpha_err}. The displayed contours are from the \lofar\ 145 MHz image of Fig.~\ref{fig:4panels} and are spaced by a factor of 2 from the 3$\sigma$ level.}
  \label{fig:3panels_alpha}
\end{figure*}

\begin{figure}
  \centering
  \includegraphics[width=\hsize,trim={0cm 0cm 0cm 0cm},clip,valign=c]{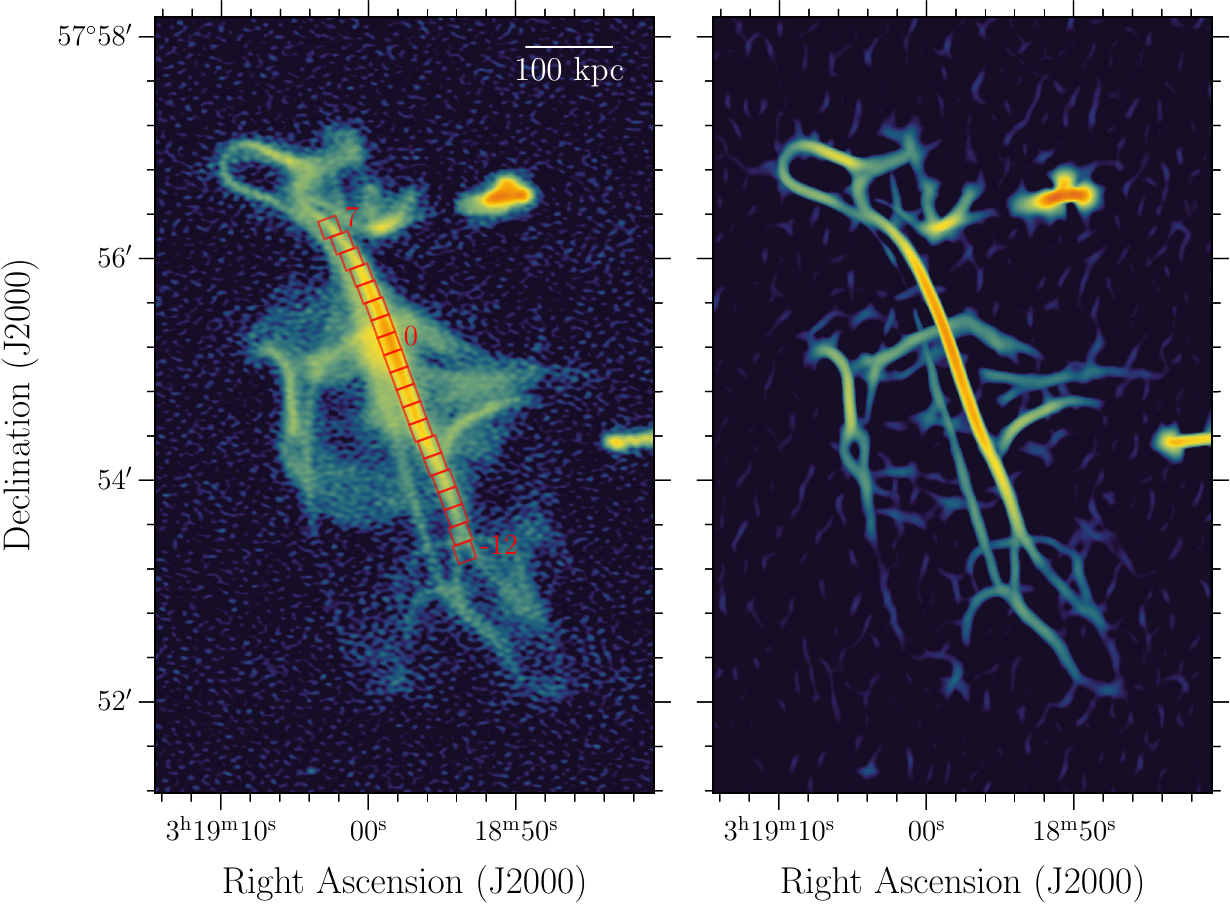}
  \caption{Zoom on the \treble. \textit{Left}: \lofar\ high-resolution image with superimposed the regions used to investigate the properties of the main filament. \textit{Right}: Sato-filtered image to emphasize the network of thin filaments in the source.}
  \label{fig:2panels}
\end{figure}

\begin{figure}
  \centering
  \includegraphics[width=\hsize,trim={0cm 0cm 0cm 0cm},clip,valign=c]{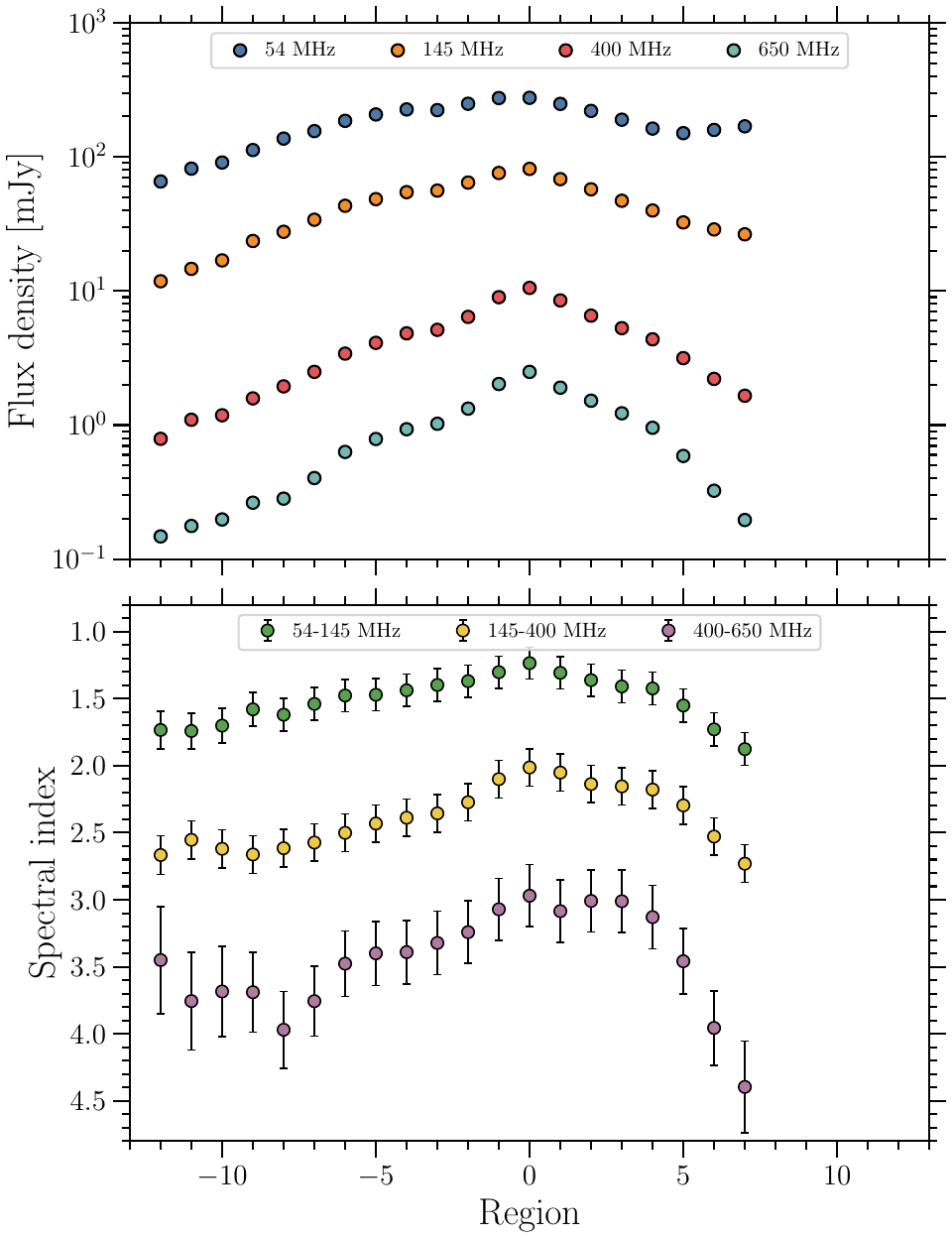}
  \caption{Flux density (\textit{top}) and spectral index (\textit{bottom}) trends for the main filament. Error bars on flux densities are smaller than the marker size. The physical separation between each region is $\approx$21 kpc.}
  \label{fig:trends}
\end{figure}

The \treble\ (Fig.~\ref{fig:lofar_highres}) has a projected largest linear size of $\approx$760 kpc and is characterized by a main bright filament with a fainter parallel counterpart to its east, both extending approximately in the north--south direction. To the north, these filaments terminate in a peculiar ``U'' morphology, while to the south they connect to a more complex network of filaments and loops. The width of the filaments in the \treble\ is unresolved at the resolution of our observations, implying transverse sizes smaller than $\approx$8 kpc. In the central region of the source, where the emission from the main radio filament peaks, the source appears more diffuse and rounded. This region is bounded on the east by a shorter filament (extending for $\approx$180 kpc in projection) with a convex morphology, and by a ``funnel''-shaped structure tapering toward the west. In addition to the \treble\, three other resolved radio sources can be identified in the field, two to the west (RG1 and RG2) and one to the east (RG3). These sources are also labeled in Fig.~\ref{fig:lofar_highres}. \\
\indent
The relative position of the radio emission with respect to the X-ray and optical images is shown in Fig.~\ref{fig:overlays}. These indicate that the filamentary source is offset with respect to the X-ray cluster center and that it does not have an obvious optical counterpart, although we note a possible connection between the lower part of the funnel structure and the BCG. Clear optical counterparts can instead be identified for the two other resolved radio sources to the west, namely WISEA J031843.95+575419.8 for RG1 and WISEA J031849.97+575632.3 for RG2. Their NIR colors, which are similar to that of the BCG, together with their radio morphologies, indicate that they are tailed radio galaxies belonging to the cluster. The other resolved source to the east of the \treble, RG3, located at the edge of the FoV shown in Fig.~\ref{fig:lofar_highres}, is likely associated with the galaxy WISEA J031929.77+575538.5. However, we cannot confidently associate this galaxy with the cluster because its emission is partially blended with that of a foreground star, although its radio morphology is also consistent with that of a tailed radio galaxy in a cluster environment. \\
\indent
In Fig.~\ref{fig:4panels} we report a multi-frequency radio view of the \treble\ from 54 to 650 MHz at a resolution of 10 arcsec, while Fig.~\ref{fig:3panels_alpha} shows the corresponding spectral index and curvature maps. The highest S/N detection is obtained from the \lofar\ 145 MHz image, where the flux density measured within the 3$\sigma$ contour is $S_{145} = 3985 \pm 399$ mJy. In the same region, the flux densities at 54, 400 and 650 MHz are $S_{54} = 24214 \pm 1457$ mJy, $S_{400} = 316 \pm 32$ mJy, and $S_{650} = 71 \pm 4$ mJy, respectively. The emission is clearly ultra-steep, with integrated spectral index values of $\alpha_{54}^{145} \approx 1.8$ at low-frequency and of $\alpha_{400}^{650} \approx 3$ at high-frequency, indicating strong spectral curvature (higher than the value of +0.5 expected in continuous injection models). The steepest regions of the source are its northern and southern tips, where we measure $\alpha_{400}^{650} > 4$. Because of its filamentary morphology and ultra-steep and curved spectrum, we classify the \treble\ as a radio phoenix. \\
\indent
In the two highest-frequency images obtained with the \ugmrt\ (Fig.~\ref{fig:4panels}), the radio phoenix appears less extended as a consequence of its steep spectral index, while a fainter diffuse emission embedding the \treble\ emerges. This fainter emission elongates for $\approx$600--700 kpc in the east-west direction and is detected at $\approx$3.2$\sigma$ at 400 MHz and at $\approx$2.3$\sigma$ at 650 MHz, based on measurements of its average surface brightness in regions selected to minimize contamination from the radio phoenix and compact sources. We note that deriving an integrated flux density for this diffuse component is challenging because of its low significance and spatial overlap with the much brighter \treble, making the separation of the two emissions uncertain. Embedded compact sources provide an additional source of contamination. Any integrated flux density would therefore strongly depend on the adopted separation between these components. For this reason, we do not report an integrated flux density and instead base our estimates on the average surface brightness of the diffuse emission in regions selected to minimize such contamination. From this, we estimate a spectral index of $\alpha_{400}^{650} = 1.2 \pm 0.4$. Instead, by adopting the average surface brightness at 400 MHz, we obtain an upper limit of $\alpha_{145}^{400} \lesssim 1.6$ ($\alpha_{145}^{400} \lesssim 2$) based on the non-detection of the emission at $2\sigma$ ($3\sigma$) level in the \lofar\ HBA image (LBA provides a shallower constraint). Given its spectral index and morphology, elongated along the same direction as the X-ray emission, the most likely interpretation of this diffuse radio component is that of a radio halo. This interpretation would be consistent with the dynamically disturbed state of the cluster, as these kinds of sources are more commonly identified in merging systems \citep[\eg,][]{cassano23}. Nonetheless, its low detection significance and the difficulty in disentangling its emission from that of the \treble\ prevent a more detailed investigation of its properties, leading us to cautiously classify it as a candidate radio halo. \\
\indent
In Fig.~\ref{fig:4panels}, the tailed radio galaxy RG1 extends in projection for $\approx$170 kpc toward the west and then bends northward with a lower surface brightness extension spanning an additional $\approx$260 kpc. The progressive steepening of the spectral index with distance suggests that this emission represents the continuation of the radio tail, while its arc-shaped morphology may indicate either the direction of the galaxy orbit or a deflection caused by the interaction with the ICM \citep[see, \eg,][for recent similar examples]{botteon21a1775, giacintucci22, lee23a514, lusetti24tails, bushi25}. \\
\indent
In the following, we focus on the interpretation of the \treble.

\begin{figure}
  \centering
  \includegraphics[width=\hsize,trim={0cm 0cm 0cm 0cm},clip,valign=c]{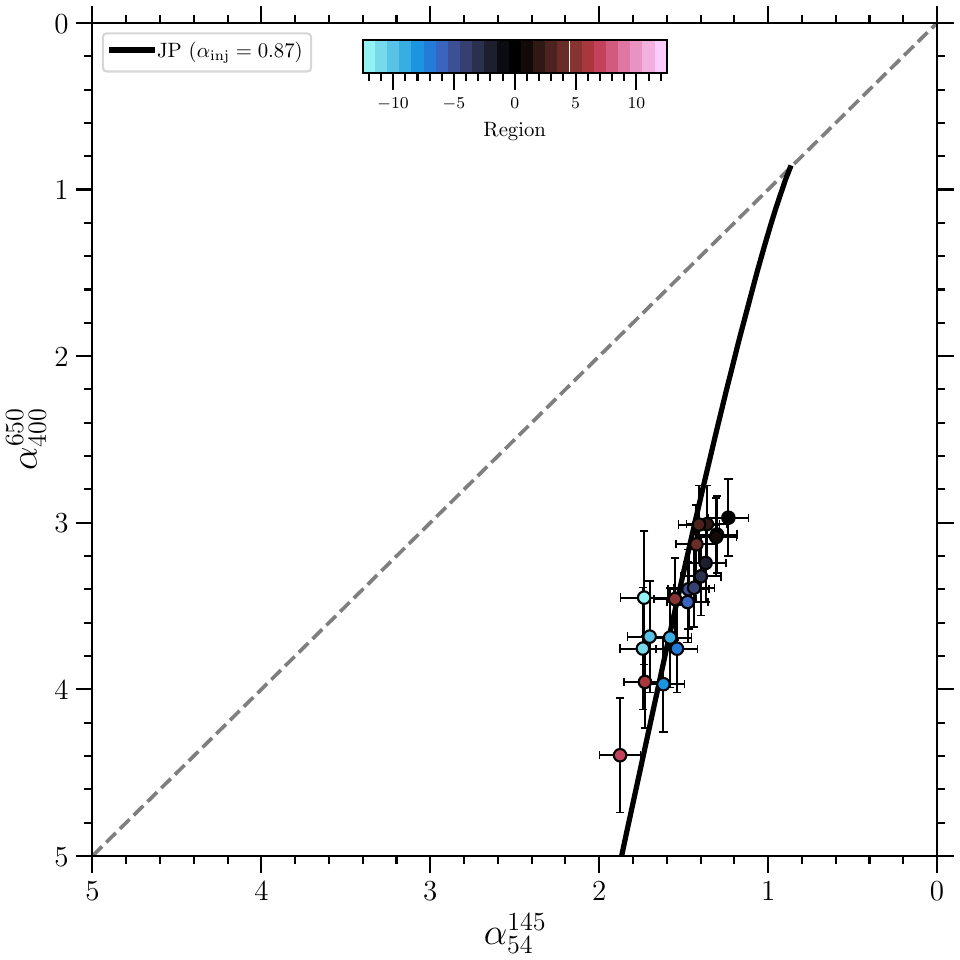}
  \caption{Color-color plot of the main filament. The black line shows the JP aging track for $\alpha_{\rm inj} = 0.87$. The black dashed line shows the 1:1 relation, equivalent to the power-law case. Data points are color-coded based on the region number.}
      \label{fig:cc}
\end{figure}

\section{Interpretation}

\subsection{Main filament trends}

\begin{figure}
  \centering
  \includegraphics[width=\hsize,trim={0cm 0cm 0cm 0cm},clip,valign=c]{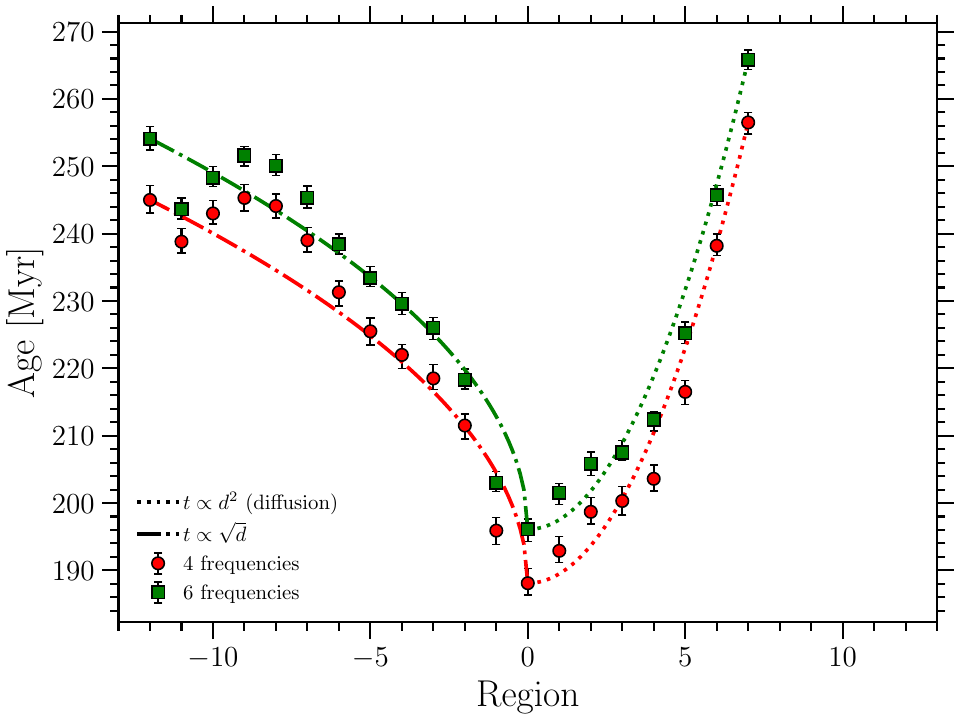}
  \caption{Radiative age of the main filament as function of region number. The dotted and dot-dashed lines indicate different scalings of the age with region number (see legend), obtained by forcing the corresponding functional forms to pass through region 0 and the outermost region of the filament. The physical separation between each region is $\approx$21 kpc.}
  \label{fig:age}
\end{figure}

Radio phoenixes are known to exhibit amorphous morphologies rich in filamentary structures ranging from a few kpc to several hundred kpc in size, and this picture is becoming increasingly clear with the progressive improvement of radio observations \citep[\eg,][]{raja24a85}. In Fig.~\ref{fig:2panels} we report a close-up view of the \treble\ to highlight its substructures, alongside an image processed with the Sato filter \citep{sato98} to better emphasize the network of thin filaments within the source. In this section, we focus on the main filament of the phoenix by analyzing the properties of the emission in the 20 square regions shown in red in Fig.~\ref{fig:2panels} (left panel). These regions have dimensions of 10 arcsec $\times$ 10 arcsec (corresponding to $\approx$21 kpc $\times$ 21 kpc) and cover the extent over which the main filament is detected above the 3$\sigma$ level in all the images of Fig.~\ref{fig:4panels}, which were used for the subsequent analysis. \\
\indent
Fig.~\ref{fig:trends} shows the flux density and spectral index trends along the main filament, where region 0 marks the position of the emission peak. The regions are numbered with positive indices toward the north and with negative indices toward the south. These plots quantitatively demonstrate the declining surface brightness and increasing spectral steepening from the center of the source moving outward. These plots also indicate that the trends are not symmetric in the two directions, with stronger gradients toward the south. The observed spectral index distribution is qualitatively similar to that seen in FRI radio galaxies and remnant radio sources, where the spectra are flatter near the location of ongoing or past particle acceleration and progressively steepen because of radiative losses as relativistic electrons are transported away from that region \citep[\eg,][]{parma07, murgia11}. In the radio phoenix scenario, this behavior may indicate that the region 0 of the \treble\ could trace the original location of the host galaxy from which the relativistic plasma was ejected before being revived. In this scenario, the fossil plasma may have retained part of the spectral gradient imprinted during its previous evolutionary stage as a radio galaxy. \\
\indent
CR electron aging models predict an increasing spectral curvature with time as a consequence of radiative losses due to synchrotron and inverse Compton emission. By constructing color-color diagrams \citep{katzstone93, rudnick94}, in which low- and high-frequency spectral indices are plotted against each other, it is possible to compare the observed behavior of a radio source with the predictions of aging models. Fig.~\ref{fig:cc} shows the color-color diagram for the main filament, highlighting once more the strongly curved spectrum, together with the evolutionary track predicted by the \citet[][JP]{jaffe73} model for an injection index of $\alpha_{\rm inj} = 0.87$ (see below). We adopt $B = B_{\rm cmb} / \sqrt{3} = 3.25 (1+z)^2 / \sqrt{3} \approx 2.3$ \muG, corresponding to the magnetic field strength that maximizes the lifetime of particles emitting synchrotron radiation at a given frequency (we remind that $t \propto \frac{B^{0.5}}{B^{2}+B_{\rm cmb}^{2}}$ and $B_{\rm cmb}$ is the equivalent magnetic field strength of the cosmic microwave background). This model (black line in Fig.~\ref{fig:cc}) provides a reasonable description of the data points, indicating that the CR electrons have the same spectral shape. \\
\indent
We exploited the broad frequency coverage in the radio band to derive the radiative age trend along the main filament by fitting the spectrum of each region with a JP model. The fitting was performed with \bratsE\ \citep[\brats;][]{harwood13,harwood15} using the four frequency data points provided by the \lofar\ observations at 54 and 145 MHz and the \ugmrt\ observations at 400 and 650 MHz. To increase the number of data points used for the spectral fitting from four to six, with the aim of better characterizing the high-frequency cutoff, we further refined the analysis by splitting \ugmrt\ band 3 and band 4 into two pairs of subbands with central frequencies of 350, 450, 600, and 700 MHz. For both the four- and six-frequency datasets, we determined the injection spectral index using the \texttt{findinject} task, obtaining $\alpha_{\rm inj}^{\rm 4\:freq} = 0.87$ and $\alpha_{\rm inj}^{\rm 6\:freq} = 0.84$. Adopting these values, we fitted a JP model assuming a constant magnetic field strength of 2.3 $\mu$G and derived the age trends reported in Fig.~\ref{fig:age}. \\
\indent
The nonthermal plasma is already old in region 0, where $t \approx 190$--$200$ Myr, and varies by only $\approx$50--70 Myr across the full extent of the filament. Given our assumed $B$, this variation corresponds to an upper limit on the age difference between the central part of the filament and its terminal ends. In the age trend plot, the asymmetric behavior between the northern and southern parts of the filament becomes even more evident. The dotted and dash-dotted lines represent quadratic and square-root scalings between the age and the region number (\ie, distance), respectively, obtained by forcing the functions to pass through the central data point and the outermost point on each side. These curves therefore do not represent best-fit functional forms to the data, although they appear to provide a good description of the observed trend. In particular, the northern part of the filament (positive indices) approximately follows a $t \propto d^2$ scaling, while the southern part (negative indices) is better described by a $t \propto \sqrt{d}$ dependence. The former scaling is expected in the case of CR diffusion, whereas the latter does not correspond to any standard propagation mode of CRs. Projection effects (\eg, the inclination of the filament along the line-of-sight) and/or different magnetic field trends along the two parts of the filament likely play a role in shaping the observed behavior. 

\begin{figure}
  \centering
  \includegraphics[width=\hsize,trim={0cm 0cm 0cm 0cm},clip,valign=c]{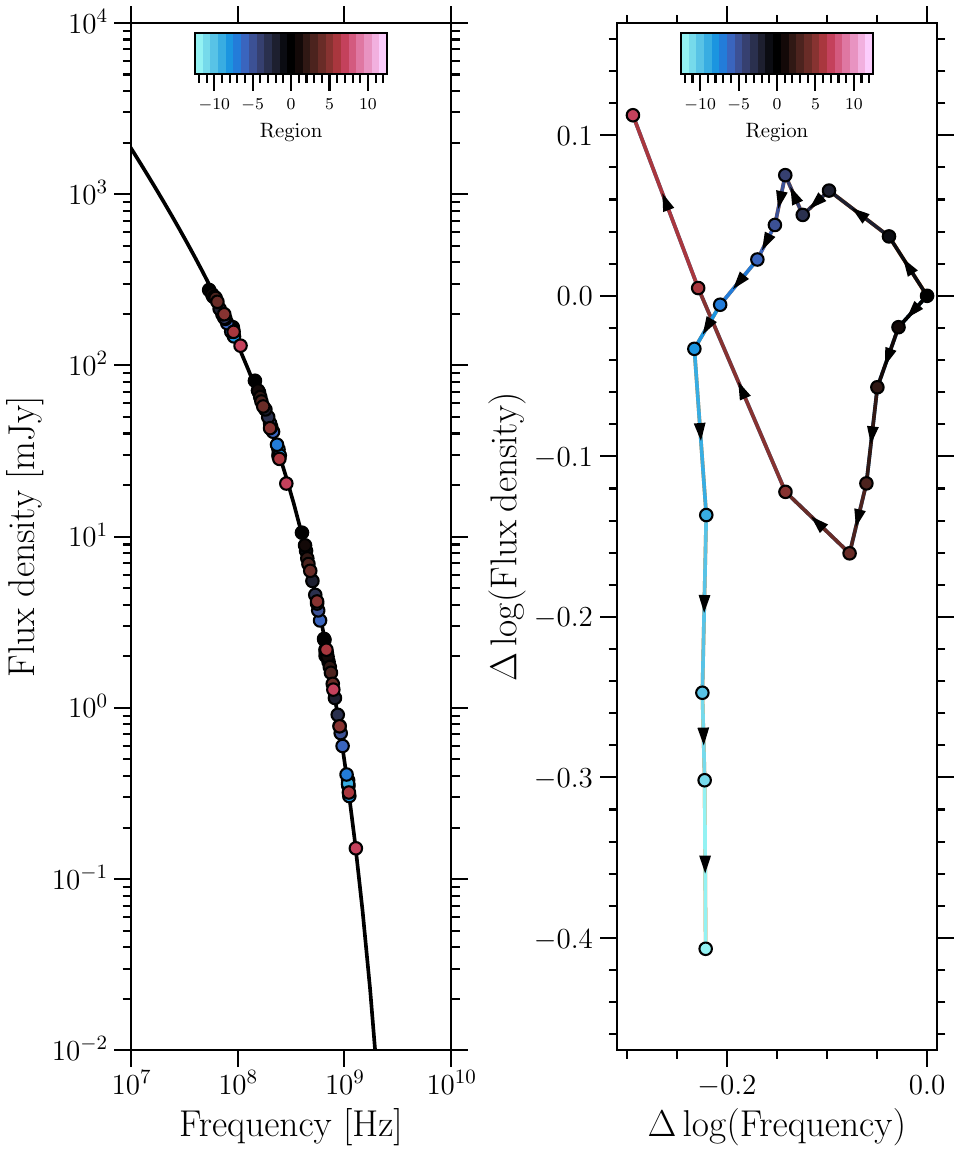}
  \caption{Global radio spectrum of the main filament (\textit{left}) with the corresponding shifts in $\log(S)$--$\log(\nu)$ space adopted to align the spectra of each individual regions to the reference one (\textit{right}). Arrows indicate the paths of the shifts, starting from region 0 (defined by $\Delta \log(S)=\Delta \log(\nu)=0$), for the northern (redder colors) and southern (bluer colors) parts of the filament.}
  \label{fig:global_shift}
\end{figure}

In order to disentangle the possible contributions of different physical effects to the observed spectrum, we applied the ``shift technique'' \citep{katzstone93,katzstone94,rudnick94} to reconstruct the global spectrum of the main filament. The underlying idea of this method is that, if the emission originates from a common population of relativistic electrons and the spectral differences between the analyzed regions are due to variations in magnetic field strength, CR electrons density, or energy, then the individual spectra can be aligned through rigid shifts in the $\log(S)$--$\log(\nu)$ plane. By applying these shifts, it is possible to recover a common global spectrum representative of the entire source \citep[see also][]{vanweeren12toothbrush}. The required shifts contain information on the physical origin of the spectral variations, with shifts in $\log(S) \propto N_{\rm T} B$ and in $\log(\nu) \propto \gamma^2 B$, where $N_{\rm T}$ is the number of CR electrons along the line-of-sight, $\gamma$ is their Lorentz factor, and $B$ is the magnetic field strength. \\
\indent
In practice, we used the spectrum of region 0 as reference (\ie, the JP model with $\alpha_{\rm inj}=0.87$ shown in Fig.~\ref{fig:cc}) and shifted the spectra of the other regions to align them with this model. The results are shown in Fig.~\ref{fig:global_shift}. The reconstructed global spectrum (left panel) indicates that the emitting regions share a similar underlying spectral shape, consistent with emission from a common electron population whose evolution has been modulated by different local conditions and timescales. At the same time, the shift diagram (right panel) shows that the spectral differences cannot be simply explained by the effect of radiative aging alone. In a pure aging scenario, the shifts in flux density and frequency are expected to follow a straight line in the $\Delta \log(S)$--$\Delta \log(\nu)$ plane, with a slope equal to $\alpha_{\rm inj}$, as observed in some passively evolving remnant radio galaxies \citep[\eg,][]{shulevski17}. By contrast, the expected slopes for variations driven solely by $B$, $\gamma$, and $N_{\rm T}$ are 1, 0, and $\infty$, respectively. Our data exhibit a complex, nonmonotonic behavior, which also differs between the northern (redder colors) and southern (bluer colors) parts of the filament. This behavior is inconsistent with a simple scenario dominated by a single varying physical parameter, including the case where the spectral steepening is exclusively driven by magnetic field variations, and suggests that multiple effects contribute to shaping the observed spectra. This, in turn, complicates a straightforward interpretation of Fig.~\ref{fig:age}. Furthermore, the observed spectral properties may be influenced by \alfvenic\ transport and mixing of relativistic particles along magnetic field filaments. If these filaments correspond to regions of depleted thermal gas, as possibly suggested by the X-ray residuals (Fig.~\ref{fig:residuals}), they may host internally low-$\beta_{\rm pl}$ plasma, enabling efficient particle propagation along the magnetic field lines \citep[\eg,][]{churazov26filaments}. \\
\indent
While the discussion above focuses on processes driven by the internal dynamics of the relativistic plasma, the interaction between the nonthermal plasma and the surrounding ICM may also affect the observed properties and morphology of the diffuse radio emission. In the following, we qualitatively discuss this alternative scenario.

\subsection{A possible formation scenario}

\begin{figure}
  \centering
  \includegraphics[width=\hsize,trim={0cm 0cm 0cm 0cm},clip,valign=c]{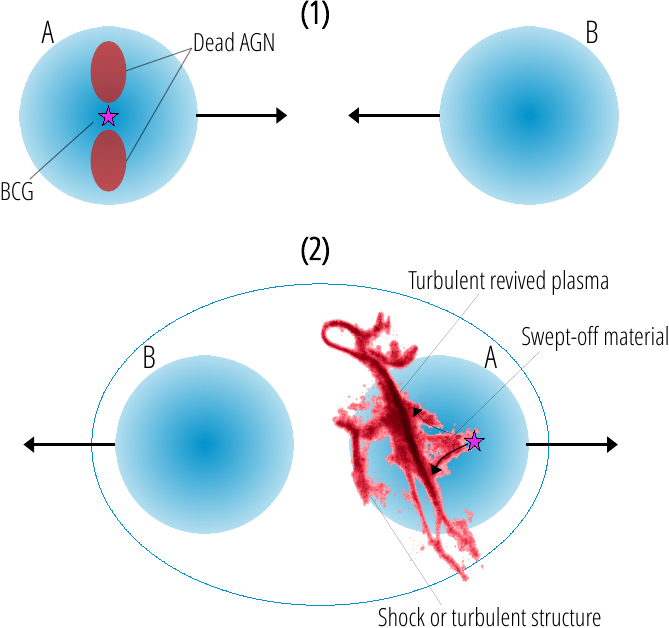}
  \caption{A possible formation scenario for the \treble. In this sketch, the configuration in which the system is currently being observed is represented by the bottom panel.}
  \label{fig:sketch}
\end{figure}

In recent years, the role of cluster radio galaxies in seeding the ICM with CR electrons and magnetic fields has become increasingly more evident \citep[\eg,][for an overview]{vazza24rev}. In particular, considerable effort has been devoted to understand the fate of the nonthermal components injected by cluster AGN, and subsequently dispersed throughout the ICM by gas motions, through the use of both idealized \citep[\eg,][]{zuhone21bubbles, dominguezfernandez24} and cosmological \citep[\eg,][]{vazza21transport, vazza23cycle} magnetohydrodynamical simulations. These studies suggest that nonthermal plasma usually develops a wide variety of morphologies under the transport by the turbulent ICM, and on the interaction between jets and the environment. Some simulations have been used to interpret, and occasionally reproduce, structures observed in radio observations of galaxy clusters that also exhibit clear signatures of interaction with the X-ray emitting thermal gas \citep[\eg,][]{hodgson21jellyfish, brienza22, lee23a514, botteon24a2657}. The morphology of the \treble\ is manifestly complex, and some analogies with the results of these numerical simulations can indeed be identified. Reproducing its morphology would require dedicated numerical simulations, with a considerable amount of uncertainties related to the description of the initial conditions, also involving a large exploration of the possible space of parameters. In what follows we instead discuss one possible scenario, that may plausibly account for the results we obtained from radio and X-ray observations. \\
\indent
We propose a simplified formation scenario in which the radio phoenix originates from aged relativistic plasma, which was initially ejected by the BCG and got subsequently revived by the cluster merger. In Fig.~\ref{fig:sketch}, we sketch two stages of a head-on collision between two clusters, A and B, in which the already aged (or ``dead'') nonthermal plasma was injected by the BCG of cluster A prior to core passage. Following core passage, corresponding to the phase in which we currently observe the system (possibly within a few $10^8$ yr after first passage, based on the disturbed X-ray morphology and the age of the radio plasma), clusters A and B move away from each other in opposite directions. As a consequence of the east-to-west motion of cluster A, the nonthermal plasma originally located near the BCG has been swept off from its host galaxy and left behind, giving rise to the ``funnel'' structure\footnote{The absence of a clear optical counterpart or the presence of significant offsets from potential hosts are typical of revived and remnant radio sources \citep[see, \eg,][for a similar case]{shulevski24}.}. In the region between the two cluster components, the gas is compressed and stretched in a direction perpendicular to the merger axis, producing the elongated morphology observed in the radio emission. At the same time, this region is expected to be turbulent, and the filaments and loops that characterize the \treble\ may trace the complex dynamics of the ICM. In this scenario, the morphology of the phoenix is shaped primarily by mixing gas motions rather than being directly associated with one or more shock surfaces, as is commonly suggested for radio phoenixes \citep{ensslin01, ensslin02relics}, although shocks likely play an important role in generating turbulence and vorticity within the medium. Whether similar processes contribute to the formation and morphology of other radio phoenixes remains to be confirmed with further observations.

\section{Conclusions}

We have reported on the observations of the \treble\ (\sou): a bright and morphologically complex radio source we first noticed by inspecting data from \lotss-DR3 \citep{shimwell26}. Although the emission had already been detected in the past, it had not been investigated for a long time \citep{green04}, and neither its nature, resolved structure, nor the environment in which it resides had previously been established. Thanks to dedicated radio and X-ray follow-up observations, together with the analysis of NIR data, we conclusively demonstrated that this source is associated with a dynamically disturbed galaxy cluster heavily obscured by Galactic absorption. From the X-ray spectroscopic analysis, we inferred a cluster redshift, temperature, and mass of $z = 0.117 \pm 0.008$, $kT = 6.0 \pm 0.6$ and $M_{500}^{\rm T-based} \approx 6.0 \times 10^{14}$ \msun, respectively. The properties of the radio emission are consistent with those of radio phoenixes, a class of sources interpreted as tracers of aged AGN remnant nonthermal plasma in galaxy clusters that has been revived and shaped by turbulent gas motions in the ICM. The faint and diffuse emission in which the \treble\ is embedded, detected solely with the \ugmrt, is a candidate radio halo generated during the ongoing cluster merger. \\
\indent
This work demonstrates the capability of low-frequency radio observations to identify galaxy clusters in the Galactic plane, whose detection at optical and X-ray wavelengths is strongly hindered by Galactic absorption \citep[\eg,][]{kollatschny21, edler26pevatron}. The \treble\ also adds to the growing class of filamentary radio sources in the ICM, which probe the complex interplay between thermal and nonthermal components in galaxy clusters. Owing to its extremely steep spectral indices, reaching values as steep as $\alpha_{400}^{650} > 4$, this source also represents one of the steepest-spectrum diffuse radio emission detected in the ICM to date \citep[see, \eg,][for other notable cases]{degasperin17gentle, ignesti20a2626, hodgson21jellyfish, edler22, raja24a85, giacintucci25}. \\
\indent
These results further highlight the importance of high-resolution low-frequency radio observations for unveiling the faint and complex nonthermal phenomenology in galaxy clusters. Future surveys and follow-up studies with \lofar2.0\ will provide unprecedented sensitivity and imaging fidelity, enabling the discovery and characterization of increasingly faint and filamentary synchrotron structures in the ICM.

\begin{acknowledgements}
We thank the anonymous referee for the comments aimed to improve the presentation of the manuscript.
We thank Walter Boschin and Arvind S.~Rajpurohit for useful discussion on optical and NIR observations.
AB thanks the Smithsonian Astrophysical Observatory for the hospitality and support during his visit where part of the work presented in this paper was carried out. 
MB acknowledges financial support from Next Generation EU funds within the National Recovery and Resilience Plan (PNRR), Mission 4 - Education and Research, Component 2 - From Research to Business (M4C2), Investment Line 3.1 - Strengthening and creation of Research Infrastructures, Project IR0000034 – “STILES - Strengthening the Italian Leadership in ELT and SKA” and from INAF under the Mini Grant 2023 funding scheme (project ‘Low radio frequencies as a probe of AGN jet feedback at low and high redshift’).
KR acknowledge support from NASA through the \xmm\ grant 80NSSC24K1852. 
IK was supported by the Simons Foundation via the Simons Investigator Award to A. A. Schekochihin.
FdG acknowledges support from the ERC Consolidator Grant ULU 101086378. 
EDR acknowledges support by the Deutsche Forschungsgemeinschaft (DFG).
FV acknowledges funding under the European Union’s Horizon Europe program through the ERC Synergy Grant COSMOMAG (Project Id. 101224803).
LOFAR \citep{vanhaarlem13} is the Low Frequency Array designed and constructed by ASTRON. It has observing, data processing, and data storage facilities in several countries, which are owned by various parties (each with their own funding sources), and which are collectively operated by the LOFAR ERIC under a joint scientific policy. The LOFAR resources have benefited from the following recent major funding sources: CNRS-INSU, Observatoire de Paris and Université d'Orléans, France; BMFTR, MKW-NRW, MPG, Germany; Science Foundation Ireland (SFI), Department of Business, Enterprise and Innovation (DBEI), Ireland; NWO, The Netherlands; The Science and Technology Facilities Council, UK; Ministry of Science and Higher Education, Poland; The Istituto Nazionale di Astrofisica (INAF), Italy.
This research made use of the Dutch national e-infrastructure with support of the SURF Cooperative (e-infra 180169) and the LOFAR e-infra group. The Jülich LOFAR Long Term Archive and the German LOFAR network are both coordinated and operated by the Jülich Supercomputing Centre (JSC), and computing resources on the supercomputer JUWELS at JSC were provided by the Gauss Centre for Supercomputing e.V. (grant CHTB00) through the John von Neumann Institute for Computing (NIC).
This research made use of the University of Hertfordshire high-performance computing facility and the LOFAR-UK computing facility located at the University of Hertfordshire and supported by STFC [ST/P000096/1], and of the Italian LOFAR-IT computing infrastructure supported and operated by INAF, including the resources within the PLEIADI special ``LOFAR'' project by USC-C of INAF, and by the Physics Department of Turin university (under an agreement with Consorzio Interuniversitario per la Fisica Spaziale) at the C3S Supercomputing Centre, Italy.
This research is part of the project LOFAR Data Valorization (LDV) [project numbers 2020.031, 2022.033, and 2024.047] of the research programme Computing Time on National Computer Facilities using SPIDER that is (co-)funded by the Dutch Research Council (NWO), hosted by SURF through the call for proposals of Computing Time on National Computer Facilities.
We thank the staff of the GMRT for support. GMRT is run by the National Centre for Radio Astrophysics of the Tata Institute of Fundamental Research.
This research has made use of data obtained from the \chandra\ Data Archive provided by the \chandra\ X-ray Center (CXC). 
In this work, observations with the eROSITA telescope onboard SRG space observatory were used. The SRG observatory was built by Roskosmos in the interests of the Russian Academy of Sciences represented by its Space Research Institute (IKI) in the framework of the Russian Federal Space Program, with the participation of the Deutsches Zentrum für Luft- und Raumfahrt (DLR). The eROSITA X-ray telescope was built by a consortium of German Institutes led by MPE, and supported by DLR. The SRG spacecraft was designed, built, launched, and operated by the Lavochkin Association and its subcontractors. The science data are downlinked via the Deep Space Network Antennae in Bear Lakes, Ussurijsk, and Baikonur, funded by Roskosmos. The eROSITA data used in this work were converted to calibrated event lists using the eSASS software system developed by the German eROSITA Consortium and analysed using proprietary data reduction software developed by the Russian eROSITA Consortium. 
This work made use of the following \textsc{python} packages: \texttt{astropy} \citep{astropy22}, \texttt{CMasher} \citep{vandervelden20}, \texttt{matplotlib} \citep{hunter07}, \texttt{numpy} \citep{vanderwalt11}, and \texttt{scipy} \citep{virtanen20}.
\end{acknowledgements}

\bibliographystyle{aa}
\bibliography{library.bib}

\begin{appendix}

\section{Impacts of ACIS calibration, spectral extraction and fitting choices}\label{app:ftemp}

\begin{figure}[h]
\centering
\includegraphics[width=0.9\columnwidth,bb=20 220 560 740]{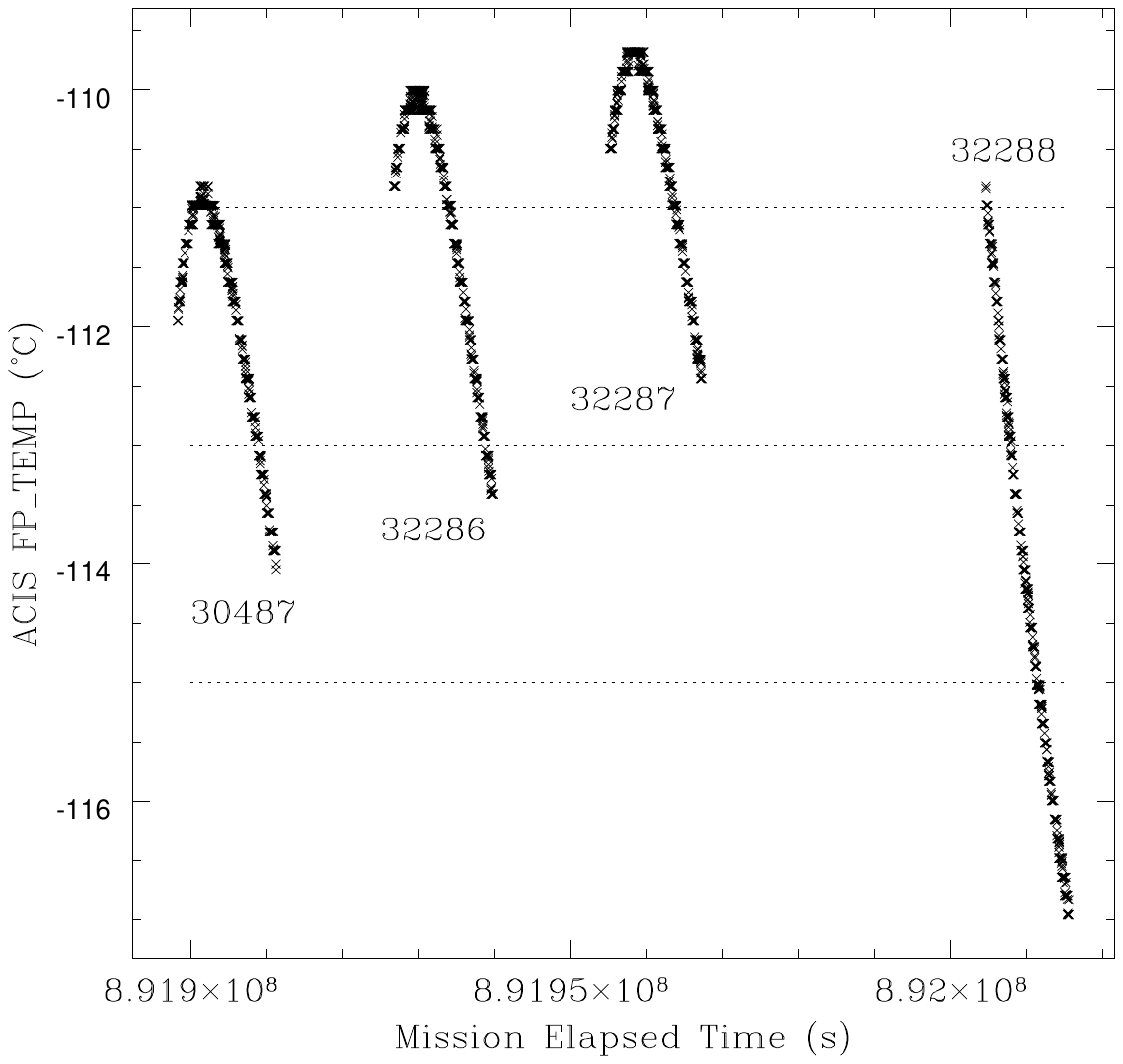}
\caption{\textit{Chandra} ACIS focal plane temperature FP\_TEMP during our observations, with ObsIDs labeled. Horizontal dotted lined indicate the bounds of the calibration temperature bins at -111, -113, -115\deg C.}
\label{fig:FP_TEMP}
\end{figure}

During our \textit{Chandra} observations the focal plane temperature of the ACIS detector (FP\_TEMP) varied significantly and, unusually, the temperature exceeded the planned limit (-111\deg C) for a large fraction ($\sim$45\%) of the total exposure. Fig.~\ref{fig:FP_TEMP} shows the FP\_TEMP during our observations, for which the planning limit was -111\deg C. \textit{Chandra} CALDB v4.12.0 introduced focal plane temperature-dependent gain calibration files for the ACIS-I CCDs. These are intended to provide improved calibration of the spectral responses for datasets such as ours. Calibration products are provided in 2\deg C bins between -105\deg C and -119\deg C, with the -119 to -120\deg C calibration unchanged from previous releases. 

To assess the potential impact of FP\_TEMP variations and spectral fitting approaches, we extracted spectra in 2\deg C bins, following the method described in the \textit{Chandra} threads\footnote{\url{https://cxc.harvard.edu/ciao/threads/acis-fptemp-resp/index.html}}. We used a large elliptical region centered on the cluster, with semi-axes of $\sim$3.6 arcmin and $\sim$2.2 arcmin, with a background extracted from a 1.25 arcmin-wide annulus close to the edge of the ACIS-I FoV. The ellipse region contained $\sim$12500 net counts (0.5--7 keV) before filtering. A number of variations on the spectra were created: 

\begin{enumerate}
    \item Filtering the data for background flaring using the \texttt{lc\_clean} algorithm, using a simple 3$\sigma$ clip, or with no filtering.
    \item Fitting individual spectra in parallel, using the \texttt{combine\_spectra} task to combine them into single spectra for each 2\deg C calibration bin, or combining all ``cool'' (FP\_TEMP$<$-111\deg C) spectra.
    \item Binning spectra to 20 counts per bin and fitting with $\chi^2$ statistics or binning to 1 count per bin and fitting with the C-statistic.
\end{enumerate}

\noindent
For each iteration, we extracted 12 individual spectra, including 4 from “hot” (FP\_TEMP$>$-111\deg C) periods. All fitting was performed as described in Section~\ref{sec:xray}. \\
\indent
Comparison of the results for the various permutations of spectra suggests that although filtering, combination and fitting choices do impact fit parameters, for the ``cool'' (FP\_TEMP$<$-111\deg C) data, the variation is within the 1$\sigma$ uncertainty bounds. We therefore conclude that our \textit{Chandra} results are robust. The ``hot'' data also generally agree to within the uncertainties, but we find systematically lower abundances and redshifts from the ``hot'' data and from spectra combining both ``hot'' and ``cool'' data. Examination of the fit statistic distribution for the redshift reveals that including the ``hot'' data broadens the distribution and produces a double-peaked minimum. This suggests that there is a noticeable gain difference between the ``hot'' and ``cool'' data, at least around the 6.7~keV FeXXV line which drives the redshift fit, and that the degree of line broadening may be greater in the ``hot'' data, with less accurate modeling of the line leading to lower apparent abundances. We therefore affirm the approach recommended in the \textit{Chandra} threads, excluding times where FP\_TEMP exceeds the planning limit for the observation.

\begin{figure*}
\centering
\includegraphics[width=0.49\hsize]{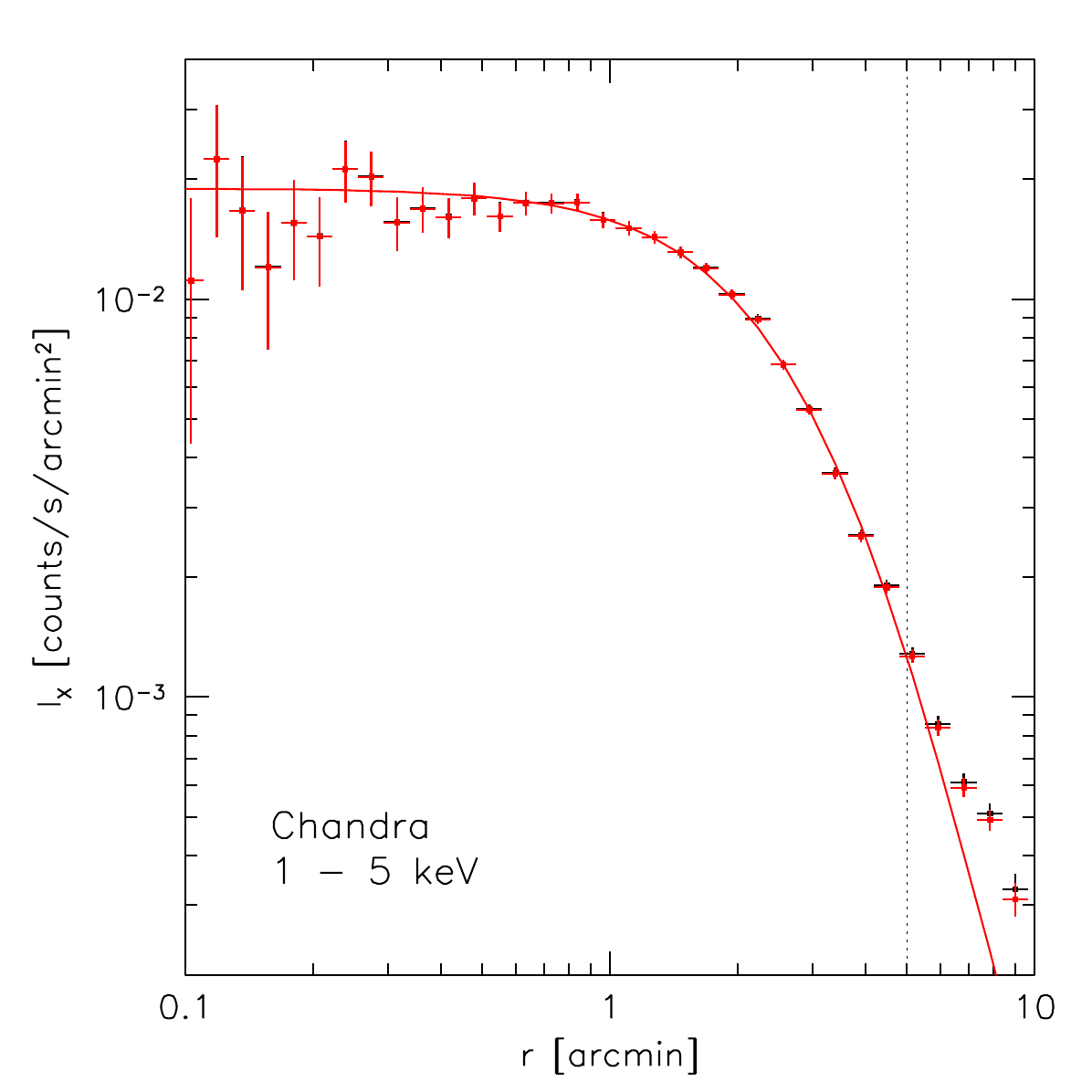}
\includegraphics[width=0.49\hsize]{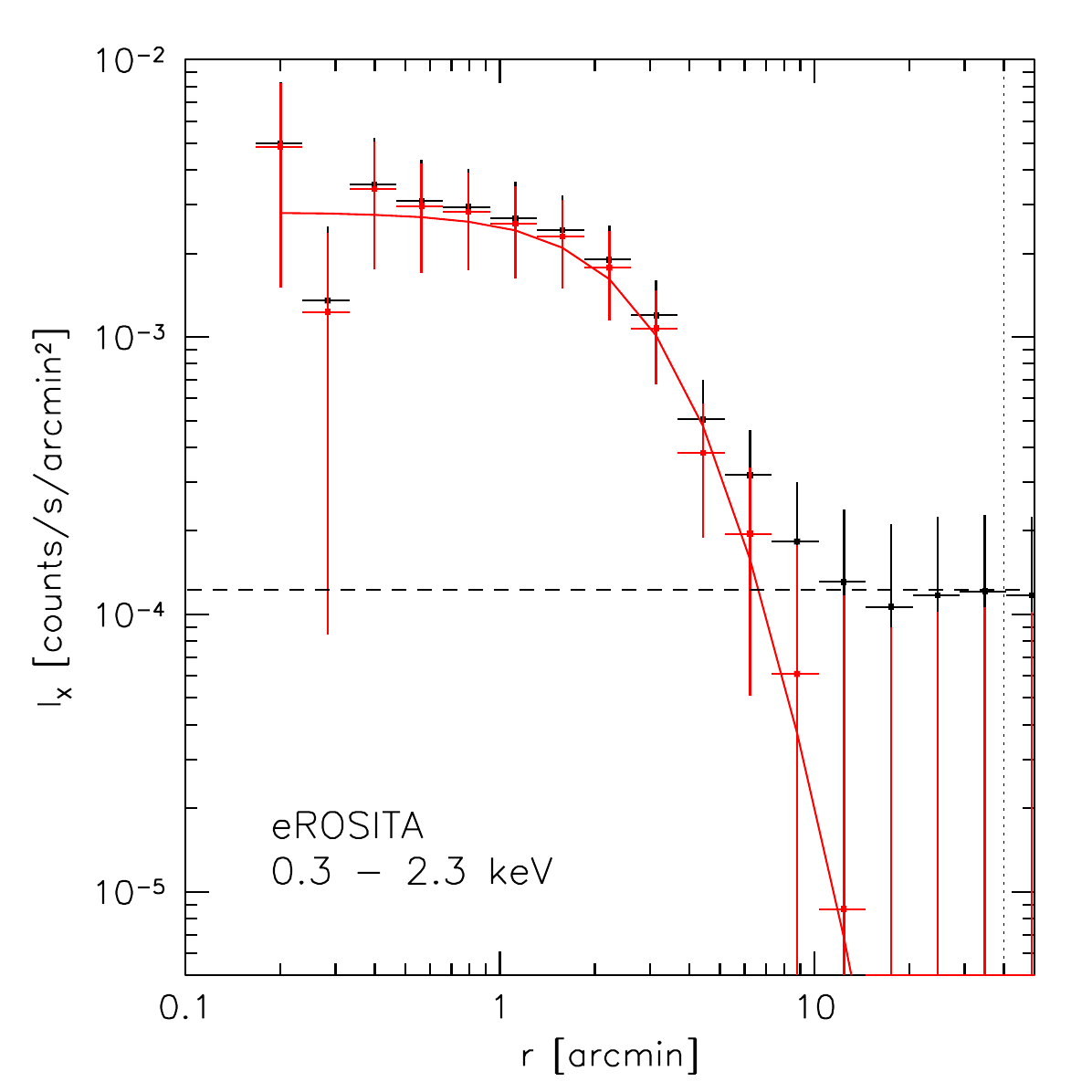}
\caption{Surface brightness profiles (black data points) measured with \textit{Chandra} (\textit{left}) and SRG/eROSITA (\textit{right}), extracted in concentric annuli centered at RA = 49.782$^{\circ}$ and Dec = 57.920$^{\circ}$. The profiles are fitted $\beta$-model (red curve) plus a constant background. The red data points show the surface brightness after subtraction of the constant background component.}
\label{fig:sb_fit}
\end{figure*}

\section{X-ray surface brightness profile}\label{app:beta}

We extracted X-ray surface brightness profiles from the \textit{Chandra} and SRG/eROSITA observations in the 1--5 keV and 0.3--2.3 keV energy bands, respectively, adopting concentric annuli centered at RA $=49.782^\circ$ and Dec $=57.920^\circ$ and fitting the resulting profiles with a $\beta$-model plus a constant background component. For the eROSITA data, the background dominates the surface brightness at radii $r \gtrsim 10'$ (see Fig.~\ref{fig:sb_fit}). The \textit{Chandra} observation covers only the central part of the cluster and does not extend to radii where the background can be measured directly. \\
\indent
The best-fit core radius is $r_{\rm c}=3.76$ arcmin for \textit{Chandra} and $r_{\rm c}=4.98$ arcmin for eROSITA, while the corresponding slope parameters are $\beta=1.06$ and $\beta=1.19$, respectively. Although the best-fit parameters differ, they provide adequately good descriptions of the same underlying surface brightness profile. This is a consequence of the well-known degeneracy between the $\beta$ and $r_{\rm c}$ parameters.

\section{Spectral index and spectral curvature error maps}\label{app:radio_err}

\begin{figure*}
  \centering
  \includegraphics[width=\hsize,trim={0cm 0cm 0cm 0cm},clip,valign=c]{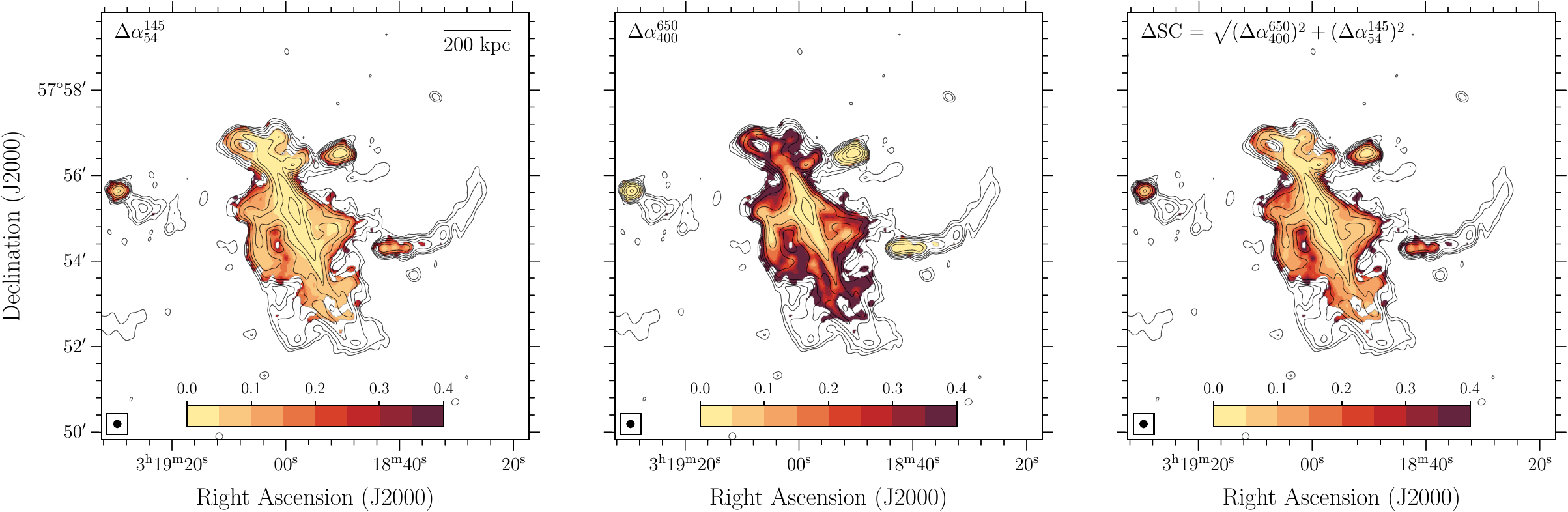}
  \caption{Images show the spectral index error maps at low (54--145 MHz, \textit{left}) and high (400--650 MHz, \textit{center}) frequency, and the corresponding spectral curvature error map (\textit{right}) associated with Fig.~\ref{fig:3panels_alpha}.}
  \label{fig:3panels_alpha_err}
\end{figure*}

\end{appendix}

\end{document}